# PRIMARY MECHANISM OF THE THERMAL DECOMPOSITION OF TRICYCLODECANE

*Olivier HERBINET, Baptiste SIRJEAN, Roda BOUNACEUR, René FOURNET, Frédérique BATTIN-LECLERC, Gérard SCACCHI, Paul-Marie MARQUAIRE**

Département de Chimie Physique des Réactions, UMR 7630 CNRS, INPL-ENSIC, 1, rue Grandville, 54001 NANCY Cedex - France

ABSTRACT

In order to better understand the thermal decomposition of polycyclanes, the pyrolysis of tricyclodecane has been studied in a jet-stirred reactor at temperatures from 848 to 933 K, for residence times between 0.5 and 6 s and at atmospheric pressure, in order to obtain a conversion between 0.01 and 25 %. The main products of the reaction are hydrogen, methane, ethylene, ethane, propene, 1,3-cyclopentadiene, cyclopentene, benzene, 1,5-hexadiene, toluene and 3-cyclopentyl-cyclopentene. A primary mechanism containing all the possible initiation steps, including those involving diradicals, as well as propagation reactions has been developed and allows experimental results to be satisfactorily modeled. The main reaction pathways of consumption of tricyclodecane and of formation of the main products have been derived from flow rate and sensitivity analyses.

* E-Mail : Paul-Marie.Marquaire@ensic.inpl-nancy.fr, Tel : +33 3 83175070, Fax : +33 3 83378120

## INTRODUCTION

The study of the thermal decomposition of cyclanes and, still more, of polycyclanes presents several kinetic difficulties which have received little attention. The unimolecular decomposition of cyclic hydrocarbons involves either the breaking of a C-H bond, which has a large bond energy (99 $kcal.mol^{-1}$) or the opening of the ring and the formation of a diradical.

Studies about gas phase reactions of polycyclanes have been the subject of several papers, but the present understanding is far from complete. The thermal decomposition and the oxidation of exo-tricyclo[5.2.1.$0^{2,6}$]decane (the structure of which is presented in Figure 1a and which is called tricyclodecane further in this paper) have been examined because it is the main component of synthetic fuels (e.g. RJ-6, JP-9 and JP-10) that are used in aircraft due to their high volumetric energy content. Its thermal decomposition have experimentally been studied in plug flow reactors with gas chromatography analyses, at temperatures up to 873 K, by Striebich and Lawrence (1) and between 903 to 968 K, by Rao and Kunzru (2), in a micro-flow tube with mass spectroscopy analysis, at temperatures up to 1700 K, by Nakra et al. (3) and, in a shock tube with analysis by UV absorption, at temperatures between 1100 to 1700 K, by Davidson et al. (4). A global mechanism has been proposed by Li et al. (5) for the oxidation of tricyclodecane and has been validated by modeling ignition delay times obtained in shock tubes (6,7). This model has also been tested by Nakra et al. (3) in order to reproduce their experimental data, but simulations did not predict the formation of several important products. That is due to the fact that the model of Li et al. (5) includes globalized reactions leading directly from the reactant to cyclopentene (8,9) and small species, such as ethylene, acetylene, propargyl and allyl radicals, without considering the formation of intermediate species.

FIGURE 1

O'Neal and Benson studied the diradical mechanism of pyrolysis of some three- and four-membered ring compounds and of some polycyclic molecules (10,11). For most species, Arrhenius parameters estimated by these authors were consistent with available experimental data. Recently Tsang (8,9) proposed reaction channels to account for the fate of diradicals in the case of cyclohexane and cyclopentane and Billaud (12) et al. and Ondruschka et al. (13) in the case of decaline, but no previous publication gives any clues about the fate of diradicals obtained from tricyclic species. Concerning propagation reactions, the rate constants of the decomposition by β-scission involving the opening of a ring or of the isomerization involving polycyclic transition states are still very uncertain.

The first purpose of this paper is to present new experimental results for the thermal decomposition of tricyclodecane performed in a jet-stirred reactor, a type of reactor different from the flow tube and shock tube used in the most part of previous studies. In order to capture the first steps of the reaction, this reactor has been operated at temperatures between 848 and 933 K corresponding to a conversion between 0.01 and 25 %. The second objective is to describe the development of an exhaustive primary mechanism including all the possible initiation channels, written by following a diradical approach as proposed by Tsang for the cyclohexane and the cyclopentane, as well as the propagation reactions deriving from the H-abstraction from tricyclodecane and 3-cyclopentyl-cyclopentene (Fig. 1b). It will be shown later in this paper that 3-cyclopentyl-cyclopentene is an important product from a kinetic point of view.

## EXPERIMENTAL STUDY OF THE THERMAL DECOMPOSITION OF TRICYCLODECANE

As the experimental apparatus used in this study has been described in details in a previous paper concerning the thermal decomposition of n-dodecane (14). The main features and the specificities of the present analyses are discussed below.

### *Experimental procedure*

Experiments were carried out in a continuous jet stirred reactor (internal volume about 90 $cm^3$) (15) made of quartz and operated at a constant temperature and pressure. The heating of the reactor was achieved by means of electrically insulated resistors directly coiled around the vessel. Temperature was measured by using a thermocouple located inside the reactor in a glass finger. In order to obtain a spatially homogeneous temperature inside the reactor, reactants were preheated at a temperature close to the reaction temperature. Corresponding residence time in the preheating section was approximately 1% of the total residence time. Experiments were performed at a pressure slightly above atmospheric pressure (106 kPa). The pressure was manually controlled by a control valve placed upstream the products analysis zone.

The liquid tricyclodecane (96.6 wt% exo form, 2.5 wt% endo form) was contained in a glass vessel pressurized with nitrogen. After each load of the vessel, nitrogen bubbling and vacuum pumping were performed in order to remove oxygen traces dissolved in the liquid hydrocarbon fuel. The liquid reactant flow rate was controlled by using a liquid mass flow controller, mixed to the carrier gas (helium, 99.995% pure) and then evaporated by passing through a single pass heat exchanger whose temperature is set above the boiling point of the mixture. Carrier gas flow rate was controlled by a gas mass flow controller located before the mixing chamber.

The analyses of the products leaving the reactor were performed in two steps due to the formation of compounds, which were either gaseous or liquid at room temperature. For the analysis of heavy species, i.e. the liquid ones at room temperature, the outlet flow was directed towards a trap maintained at liquid nitrogen temperature during a determined period of time to concentrate the sample. At the end of this period (typically between 5 and 15 min), the trap was removed and, after the addition of acetone and of an internal standard (n-octane), progressively heated up. When the temperature of the trap was close to 273 K the mixture was poured into a small bottle and then injected by an auto-sampler in a gas chromatograph with flame ionization detection (FID) for quantification. The column used for the separation was an HP-1 capillary column with helium as carrier gas (oven temperature profile was: 313 K held 30 min, 5 $K.min^{-1}$, 453 K held 62 min). In the conditions of this study, the column provided a good separation for the following products (retention time in min), cyclopentene (3.6), 1,3-cyclopentadiene (3.7), benzene (4.9), 1,5-hexadiene (4.5), toluene (7.9), exo-tricyclodecane (41,2), 3-cyclopentylcyclopentene (41.6) and endo-tricyclodecane (42.7). Identification of heavy species was performed by a gas chromatography-mass spectroscopy system working in the same conditions as the chromatograph used for the quantification. Calibrations were performed by analyzing a range of solutions containing known amounts of n-octane and of the compound quantified.

For the analysis of light species, like hydrogen and hydrocarbons containing less than five atoms of carbon, analysis was performed on-line using two gas chromatographs in parallel. The first chromatograph was fitted with both thermal conductivity detector (TCD) for hydrogen detection and FID for methane and $C_2$ hydrocarbons detection and used a carbosphere packed column. Hydrocarbons analysis was performed with an oven temperature of 473 K. In these conditions the column provided a good separation for: methane (2.4), acetylene (5.3), ethylene (7.3), ethane (9.6) and propene (40.7). The separation of the hydrogen peak from the helium peak (experimental carrier gas) was obtained with a oven temperature of 303 K. Argon was used as column carrier gas and also as reference gas for the TCD in order to have the better sensibility for hydrogen detection. The second gas chromatograph, which was used for $C_3$ and $C_4$ hydrocarbons analysis, was fitted with FID and a Haysep D packed column with nitrogen as carrier gas (oven temperature profile: 313 K held 30 min, rate $1K.min^{-1}$, 473 K held 30 min). Analyzed species and their relative retention times were: methane (2.6), ethane (15.5), propene (61.2) and cyclopentadiene (147.8). Light species identification and calibration were realized by injection of gaseous standard mixtures provided by Air Liquide.

Data obtained using the three different chromatographs were compared through mole fractions of products which were detected by two chromatographs, i.e., mole fractions of cyclopentadiene obtained using Haysep-D column were compared to those obtained using HP-1 column. The variation between the mole fractions obtained with the different chromatographs were less than 5%. A study of the reproducibility of the experimental results was carried out for both kinds of products analyses. Calculated maximum uncertainties on the experimental results were ± 5% with the online analysis (light species analysis) and ± 8% for the discontinuous process of analysis (heavy hydrocarbons analysis).

*Experimental results*

The thermal decomposition of tricyclodecane has been studied in the above-described reactor operated at atmospheric pressure, temperatures between 848 and 933 K, residence times from 0.5 to 5 s and initial hydrocarbon mole fractions from 0.7 to 4 % with dilution in helium. In all the figures presented in this part, the points refer to experimental observations and the curves come from simulations.

Figure 2 presents the evolution of the conversion of tricyclodecane with residence time at several temperatures (873, 893, 913 and 933 K) for an initial hydrocarbon mole fraction of 4 % (Figure 2a) and with temperature at a residence time of 1s and for several initial hydrocarbon mole fractions (0.7, 2, 4%) (Figure 2b). Conversions (0.01-25%) have been deduced from the sum of the concentrations of reaction products, because the values deriving from the difference between the amounts of reactant entering and leaving the reactor were not accurate enough (relative uncertainty on conversion is better than 10%). Figure 2b shows that the conversion is a function of reactant concentration, as then that the current order of reaction is different from 1. Figure 2c displays results obtained in the same conditions for n-dodecane with an initial concentration of 2 % (14). This results shows that the linear alkane is much more reactive than the tricyclane, despite the fact that the later is a strained molecule.

FIGURE 2

For experiments performed at temperatures between 873 and 933 K and residence times between 0.5 and 6 s, 11 main products of the reaction have been analyzed : hydrogen, methane, ethylene, ethane, propene, 1,3-cyclopentadiene, cyclopentene, benzene, 1,5-hexadiene, toluene and 3-cyclopentyl-cyclopentene. Figures 3 and 4 present the evolution of the mole fractions of these products with residence time at several temperatures (873, 893, 913 and 933 K) for an initial hydrocarbon mole fraction of 4 %.

FIGURES 3-4

These figures show that the major products of the thermal decomposition of tricyclodecane are hydrogen, ethylene, propene and 1,3-cyclopentadiene, which have selectivities always higher than 10 % and that the formation of cyclopentene is lower than that of cyclopentadiene and close to those of benzene and toluene.

As the differences between the masses of reactant entering and leaving the reactor were not accurately enough determined for such low conversions (lower than 25%), the mass balance could not be directly checked. Nevertheless, the ratio between the numbers of C-atoms and H-atoms present in the molecules of reaction products (reactant excluded) can be calculated whatever the conversion and be compared to the theoretical value which corresponds to the same ratio in the initial hydrocarbon (tricyclodecane) molecule. This ratio has been calculated for every experiment performed in this study and an average value of 0.61±0.03 (close to the theoretical value of 0.625) was obtained.

Primary products have been determined from an analysis of the selectivities (the selectivity of a product is the ratio between the mole fraction of this product and the sum of the mole fractions of all products -non converted reactant excluded-). A species is probably a primary product if the extrapolation to origin of its selectivity versus residence time gives a value different from zero. Selectivities of products versus residence time have been calculated for the experiments performed at 873 K (Figures 3-4), for which conversion is lower than 2%. Figure 5a displays the selectivities of 1,3-cyclopentadiene, cyclopentene, 1,5-hexadiene and 3-cyclopentyl-cyclopentene and shows that these products seem to be primary products. 1,3-cyclopentadiene is often a secondary product deriving from cyclopentene, but here it probably has a primary source. Ethylene, which is not shown in Figure 5, seems also to be a primary product. At this temperature, benzene and toluene present trends of primary products but their initial selectivities are very close from zero. We have then performed experiments at 848 K in order to plot the selectivities of benzene and toluene vs. residence time at a conversion lower than 0.5% as shown in figure 5b. That allows us to conclude to a probable primary character of these aromatic compounds.

FIGURE 5

DESCRIPTION OF THE MODEL

The detailed kinetic model has been constructed by using the same systematic way as EXGAS software, which performs an automatic generation of mechanisms, but cannot be directly applied to polycyclic compounds. EXGAS has been mainly developed and tested to model the oxidation of alkanes (16,17), but it has been recently used to produce a mechanism for the pyrolysis of n-dodecane (18). Except for diradicals for which the calculations are detailed further in the text, specific heats, standard enthalpies of formation and entropies of all the considered molecules or free radicals have been calculated using THERGAS software (19) based on the group and bond additivity methods and on statistical thermodynamics approach proposed by Benson (20). According to the CHEMKIN II formalism, the thermochemical data are stored as 14 polynomial coefficients (21).

**General structure of the mechanism**

The mechanism presented here includes two parts:

- A $C_0$-$C_6$ reaction base

This reaction base includes all the reactions involving radicals or molecules containing less than six carbon atoms ($C_0$-$C_2$ reaction base (22)), the reactions of $C_3$-$C_4$ unsaturated hydrocarbons (23), such as propene and the reactions of $C_5$ to $C_6$ hydrocarbons (24). The kinetic data used in these reaction bases were taken from the literature. This $C_0$-$C_6$ reaction base has been used here after removing all the species or reactions involving the presence of oxygen atoms.

- A comprehensive primary mechanism

Since, as it will be shown further in the text, under the studied conditions, 3-cyclopentyl-cyclopentene is the quasi exclusive initiation product from tricyclodecane, the two only molecular reactants considered in this part of the mechanism are tricyclodecane and 3-cyclopentyl-cyclopentene. The primary mechanism includes the following elementary steps:

- Unimolecular initiations by bond fission,
- Propagations by metathesis, or hydrogen abstraction, from the two reactants,
- Propagations by β-scission decomposition,
- Propagations by isomerization involving cyclic transition state,
- Terminations by combination of two free radicals; radicals involved in combination are $C_0$-$C_4$ radicals, radicals deriving directly from tricyclodecane by metatheses and stabilized free radicals that cannot break through β-scission of a $Csp^3$-$Csp^3$ or $Csp^3$-H bond.

The following parts of this paragraph describe which initiations and propagations have been written for tricyclodecane and 3-cyclopentyl-cyclopentene and how the related rate constants have been estimated.

**Unimolecular initiations by bond fission from tricyclodecane and 3-cyclopentyl-cyclopentene**

The unimolecular initiations by breaking of a C-H bond have been considered through their corresponding reverse reactions, the combinations of the two radicals with a rate constant taken from Allara and Shaw (25).

In the case of linear or ramified alkanes, the unimolecular initiations by breaking of a C-C bond lead to the formation of two free radicals, while diradicals are obtained in the case of cyclic and polycyclic hydrocarbons, as shown in Figure 6 for tricyclodecane and in Figure 7 for 3-cyclopentyl-cyclopentene.

FIGURES 6-7

*a/ Estimation of the thermodynamic properties of diradicals*

Calculations of thermodynamic data for diradicals were not directly possible using THERGAS. For a diradical (•R•), the thermodynamic data of the two related mono-radicals (HR•) and (•RH) and of the molecule (HRH, obtained by adding two hydrogen atoms to the diradical •R•) were obtained with THERGAS and corrections due to the second radical center were estimated assuming that there was no interaction between the two radical centers. The enthalpy of formation of •R• ($\Delta_f H°_{(298K)}$(•R•)) was then obtained according to equation {1}:

$$\Delta_f H°_{(298K)}(\bullet R\bullet) = E(R\text{-}H) + \Delta_f H°_{(298K)}(HR\bullet) - \Delta_f H°_{(298K)}(H\bullet) \qquad \{1\}$$

with: E(R-H) : energy of the bond broken in HR• to give •R•,

$\Delta_f H°_{(298K)}$ (HR•) : enthalpy of formation of HR•,

$\Delta_f H°_{(298K)}$ (H•) : enthalpy of formation of H• (52,1 $kcal.mol^{-1}$ (20)).

The entropy of a species S ($S°_{(298K)}$(S)) can be estimated from the entropy of a model compound M ($S°_{(298K)}$(M)) according to equation {2} established by Benson (20):

$$S°_{(298K)}(S) = S°_{(298K)}(M) + R\ln\left(\frac{\sigma_M}{\sigma_S}\right) + R\ln\left(\frac{n_S}{n_M}\right) + R\ln(2s+1) + C \qquad \{2\}$$

with: $\sigma_M$ and $\sigma_S$: the total symmetry numbers of M and S, respectively.

$n_M$ and $n_S$: the number of optical isomers of M and S, respectively.

Rln(2s+1): the electronic correction. The spin s is equal to 0 in the case of a molecule or a diradical in the singlet state, to 1/2 for a free radical and to 1 for a diradical in the triplet state. For a diradical (mixture of 75% in the triplet state and 25% in the singlet state (20)) the electronic contribution is equal to 2Rln(2). R is the gas constant.

C is the sum of the other corrections due to the differences of structure between S and M (free internal rotations, barrier corrections, vibrations of H atoms, translations and external rotations).

Applying equations {2} to the case of •R•, •RH, HR• and HRH, we obtain equation {3}:

$$S°_{(298K)}(\bullet R\bullet) = S°_{(298K)}(HR\bullet) - (S°_{(298K)}(HRH) - S°_{(298K)}(\bullet RH)) - R\ln\left(\frac{\sigma_{HRH}\sigma_{\bullet R\bullet}}{\sigma_{\bullet RH}\sigma_{HR\bullet}}\right) - R\ln\left(\frac{n_{\bullet RH}n_{HR\bullet}}{n_{HRH}n_{\bullet R\bullet}}\right) \qquad \{3\}$$

In the case of diradical formed by the initiation (1) of Figure 6, $\sigma_{RH}$=2, $\sigma_{\bullet RH}$=1, $\sigma_{HR\bullet}$=1 and $\sigma_{\bullet R \bullet}$=1, the symmetry correction is equal to –Rln(2); $n_{RH}$=1, $n_{\bullet RH}$=1, $n_{HR\bullet}$=1 and $n_{\bullet R \bullet}$=1, the optical isomers correction is equal to 0.

According to Benson (20), the specific heat of •R• ($Cp_T$(•R•)) was deduced from equation {4}:

$Cp_T$(•R•) = $Cp_T$ (HR•) - ( $Cp_T$ (HRH)- $Cp_T$ (•RH)) {4}

Table 1 presents the thermodynamic data of the 7 diradicals directly obtained from tricyclodecane and for the 2 radicals and the 5 diradicals deriving from 3-cyclopentyl-cyclopentene. In the case of 3-cyclopentyl-cyclopentene only breakings of alkylic and allylic C-C bonds were taken into account because activation energies of these reactions are much lower than the activation energy of the breakings of vinylic C-C bonds. This table shows that a good agreement is obtained between the estimations above explained and much time consuming quantum calculations in the example of diradical BR1. Thermodynamic data have been calculated from geometric parameters and vibrationnal frequencies obtained at the UB3LYP/cbsb7 level of theory (26). The diradical BR1 was computed as a singlet state within the broken symmetry spin-unrestricted approach using the GUESS=MIX option of Gaussian 03 (27). This method has been shown to provide reliable energies and geometries (28-29). Computations were performed using the Gaussian 03 suite of programs. Isodesmic reaction analysis is used to further improve accuracy on enthalpy of formation. Isodesmic reactions are actual or hypothetical reactions in which the types of bonds that are made in forming the products are the same as those which are broken in the reactant. To obtain more accurate enthalpies of formation it is better to use isogyric reactions (reactions that conserve the spin) and to choose species which own structures as similar as possible (Examples of isodesmic reactions used to estimate the enthalpy of reaction of diradical BR1 are given in Figure 8).

TABLE 1

FIGURE 8

*b/ Estimation of the kinetic parameters of the unimolecular initiations*

Kinetic parameters of the unimolecular initiations have been deduced from data from the literature (10,11). An average value of $5.0\times10^{15}$ $cm^3.mol^{-1}.s^{-1}$ has been used for the A factor, which is close to what is used by EXGAS software for acyclic alkanes (17). This value is close to A factors of numerous reactions of thermal decomposition of cycloalkanes and polycycloalkanes (20). According to O'Neal and Benson (10,11), activation energy of the opening of a (poly)cyclane is given by {5} :

$$E_1=DH(C\text{-}C)-\Delta E_{TC}+E_{-1} \qquad \{5\}$$

With $E_1$ the activation energy of the reaction of the opening of the cycle.

DH(C-C) the bond energy of the broken C-C bond.

$\Delta E_{TC}$ the difference of ring strain energy occurring through the reaction.

$E_{-1}$ the activation energy of the reaction of the closure of the diradical.

It was observed that the sum DH(C-C)+$E_{-1}$ is almost constant (it fluctuates around 87 $kcal.mol^{-1}$) for numerous cycloalkanes such as cyclohexane, cyclopentane, cyclobutane and cyclopropane. For tricyclodecane activation energy $E_1$ was obtained by substracting the difference of ring strain energy $\Delta E_{TC}$ to 87 $kcal.mol^{-1}$. When the broken bond was in β position of a double bond the activation energy was estimated in the same way by assuming that the sum DH(C-C)+$E_{-1}$ was close to the activation energy of the breaking of the allylic bond in 1-pentene (Ea=71.2 $kcal.mol^{-1}$ (8)). The ring stain energy of tricyclodecane (22 $kcal.mol^{-1}$) has been deduced from the enthalpy of formation experimentally measured by Boyd et al. (31) by subtracting the sum of the contributions of groups as proposed by Benson (20). The ring stain energy of 3-cyclopentyl-cyclopentene (12 $kcal.mol^{-1}$) is assumed as twice that of cyclopentane. The breaking of vinylic C-C bonds in 3-cyclopentyl-cyclopentene has not been taken in account because activation energy of such reaction is much higher (102 $kcal.mol^{-1}$) than the ones of breakings of alkylic and allylic C-C bonds. Table 2 summarizes the values taken for the energies of activation of the reactions of unimolecular initiation by breaking of a C-C bond from tricyclodecane and 3-cyclopentyl-cyclopentene.

TABLE 2

**Fate of the diradicals obtained through the unimolecular initiations of tricyclodecane and 3-cyclopentyl-cyclopentene**

The fate of cyclic and polycyclic diradicals has not yet been much studied. Tsang (8, 9) has proposed some channels in the case of diradicals deriving from cyclohexane and cyclopentane. Figure 9a shows how the diradical deriving from cyclohexane can either give back the initial hydrocarbon by cyclisation (reverse reaction of unimolecular initiation) or react by β-scission decompositions to give molecules and/or smaller diradicals and by isomerization to form 1-hexene. This isomerization, which involves the internal transfer of a H-atom in β-position of a radical center, transforms a diradical in a stable molecule and acts then as an internal disproportionnation. Billaud et al. (12) have proposed similar types of reactions in the case of decaline. The energy diagram of Figure 9b shows the energy levels associated to the reactions involved in cyclohexane unimolecular initiation step. Energy of the transition state appears clearly smaller in the case of the isomerization of the biradical obtained from unimolecular initiation of cyclohexane than in the case of its reaction of decomposition through β-scission.

FIGURE 9

We have then systematically written all the β-scission decompositions and internal isomerizations for the 12 diradicals directly obtained from tricyclodecane and from 3-cyclopentyl-cyclopentene and for those deriving from them by β-scission decompositions. Only the β-scission decompositions involving the breaking of a $Csp^3$-$Csp^3$ bond have been considered, except when the only possible decomposition was the breaking of a $Csp^3$-$Csp^2$ bond (activation energy 6.8 $kcal.mol^{-1}$ higher than the one of the breaking of a $Csp^3$-$Csp^3$ bond). Tables 3 and 4 show the products obtained by β-scissions and isomerizations for the diradicals formed by the initiation (1) of Figure 6 and by the initiation (2) of Figure 7. We have thus obtained 250 reactions involving 48 diradicals, 6 radicals and 127 molecules for the initiation step of tricyclodecane and 38 reactions involving 9 diradicals, 3 radicals and 23 molecules for the initiation step of 3-cyclopentyl-cyclopentene.

TABLES 3-4

*a/ Kinetic parameters of the β-scission decompositions of diradicals*

The rate constants used for the β-scission decompositions of diradicals are the same as those used for free radicals, the estimation of which will be detailed further in the text, but for the breaking of a $Csp^3$-$Csp^3$ bond in β–position of two radical centers, as shown in Figure 10. The activation energy of the β-scission of Figure 10a, which leads to the formation of two alkenes, is low and has been estimated to 2.8 kcal.mol$^{-1}$ by CBS-QB3 method (32) performed with Gaussian03 software (27). The use of CBS-QB3 method on diradical species in a broken symmetry approach involves a modification of the empirical correction included in CBS-QB3, this correction has been described elsewhere and was shown to provide reliable energies (33). In all the quantum estimations of rate parameters presented in this paper, Intrinsic Reaction Coordinate calculations have been performed to insure that the transition states connect correctly the reactants and the products. The same value was used for the activation energy of the β-scission in the case of cyclic diradicals in which a single cycle bears two radical centers (such as 1,4-cyclohexadiyl diradical) except for the particular case of 1,3-cyclopentadiyl diradical for which Ea is higher (22 kcal.mol$^{-1}$) because of steric inhibition of π bonding (20).

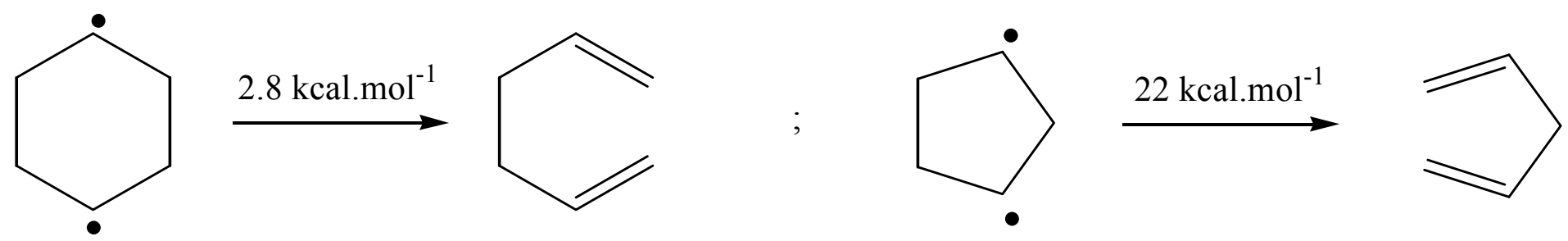

The activation energy of the β-scission of Figure 10b was estimated as that of the model reaction displayed in Figure 10c assuming that, for acyclic compounds, the activation energy of a β-scission decomposition Ea (acyclic β-scission) is a linear function of its enthalpy of reaction $\Delta_r H_{(1000K)}$. As shown in Figure 11, this has been proved correct for 12 reactions with the following Evans-Polanyi relationship (in kcal.mol$^{-1}$):

$$Ea_{(\text{acyclic } \beta\text{-scission})} = 18.9 + 0.53 \times \Delta_r H_{(1000K)} \qquad \{6\}$$

The enthalpy of the reaction of Figure 10c is 6.8 kcal.mol$^{-1}$ and then according to equation {6} the activation energy is 22.5 kcal.mol$^{-1}$.

FIGURES 10-11

*b/ Kinetic parameters of the isomerizations of diradicals*

The A-factors used for the isomerizations of diradicals are mainly based on the changes in the number of losses of internal rotations as the reactant moves to the transition state and on the reaction path degeneracy (here the number of transferable atoms of hydrogen) (17). The activation energies are deduced from that of 4 reference reactions, the isomerization of trimethylene diradical (•CH2-CH2-CH2•) to form propene ($Ea_{(ref.\ 3)}$ = 9.5 kcal.mol$^{-1}$) as proposed by Tsang (8) and the isomerizations of the diradicals deriving from cyclobutane ($Ea_{(ref.\ 4)}$ = 16.3 kcal.mol$^{-1}$), cyclopentane ($Ea_{(ref.\ 5)}$= 7.75 kcal.mol$^{-1}$) and cyclohexane ($Ea_{(ref.\ 6)}$ = 3.85 kcal.mol$^{-1}$) for which the activation energy is calculated by density function theory calculations using CBS-QB3 method (32) performed with Gaussian03 software (27). The activation energy, $Ea_{(isom.\ i)}$, of an isomerization involving the formation of a polycyclic transition state with the creation of a cycle of size i (number of members in the ring) was estimated as follow:

$$Ea_{(isom.\ i)} = Ea_{(ref.\ i)} - SE_{(cycle\ i)} + SE_{(polycycle)} \qquad \{7\}$$

with : $SE_{(cycle\ i)}$ is the strain energy of the cycle of size i used as the reference ($SE_{(cycle\ 3)}$ = 27.6 kcal.mol$^{-1}$, $SE_{(cycle\ 4)}$ = 26.2 kcal.mol$^{-1}$, $SE_{(cycle\ 5)}$ = 6 kcal.mol$^{-1}$, $SE_{(cycle\ 6)}$ = 0 kcal.mol$^{-1}$ (20)),

$SE_{(polycycle)}$ (Table 5) is the actual strain energy of the new polycycle created (34-36).

The strain energy due to the presence of side rings have been neglected.

The structures of the transition states assumed in the case of the isomerisations of Table 3 are shown in Figure 12. The isomerization of Figure 12a involves the formation of a six-membered ring embedded in a polycycle having a structure close to that of bicyclo[3.2.1]octane ($SE_{(polycycle)}$ = 16 kcal.mol$^{-1}$). The isomerization of Figure 12b involves the formation of a five-membered ring embedded in a polycycle having a structure close to that of bicyclo[2.2.1]heptane (norbornane) ($SE_{(polycycle)}$ = 16 kcal.mol$^{-1}$). As the isomerization of Figure 12c involves the formation of a five-membered ring with side rings, its activation energy is directly equal to 7.75 kcal.mol$^{-1}$. The isomerization of Figure 12d occurs through the formation of a seven-membered ring (we assume $Ea_{(ref.\ 7)}$= $Ea_{(ref.\ 5)}$) included in a polycycle having a structure close to that of tricyclo[5.2.1.1$^{2,5}$]undecane ($SE_{(polycycle)}$ taken equal to that of bicyclo[4.2.1]nonene, i.e. 26 kcal.mol$^{-1}$).

FIGURE 12

Simulations performed in the conditions of our experimental study have shown that 99% of the flux of consumption of tricyclodecane through the unimolecular initiations involving the formation of diradicals leads to the formation of 3-cyclopentyl-cyclopentene. Therefore, amongst the 127 molecules which are obtained by unimolecular initiations of the tricyclodecane, we have only written the reactions of unimolecular initiation and of metathesis involving 3-cyclopentyl-cyclopentene.

**Propagation reactions**

Propagation reactions include metatheses from the reactant, β-scission decompositions and isomerizations.

*a/ Propagations by metathesis from the reactant*

Species which have been considered for the reactions of metathesis with the tricyclodecane and the 3-cyclopentyl-cyclopentene are small free radicals (hydrogen radicals, methyl radicals and ethyl radicals) and stabilized free radicals that cannot break through β-scission of a $Csp^3$-$Csp^3$ or $Csp^3$-H bond (allyl radicals for example). The abstractions of vinylic H-atoms have been neglected. Reactions of metatheses with tricyclodecane lead to the formation of the 6 tricyclic radicals presented in Figure 13a and with 3-cyclopentyl-cyclopentene to the 6 radicals shown in Figure 13 b. The correlations used for the estimation of the rate constants have been deduced from those used by EXGAS software (16, 37, 38) and are given in Table 6.

FIGURE 13

TABLE 6

*b/ Propagations by β-scission*

As the energy of the $Csp^3$-H bond is about 13 kcal.mol$^{-1}$ higher than that of the $Csp^3$-$Csp^3$ bond, β-scission decompositions by $Csp^3$-$Csp^3$ bond fission have been considered preferentially to β-scissions by $Csp^3$-H bond fission, except when only $Csp^3$-H bond fission was possible. Table 7 displays examples of β-scission decompositions in the case of the tricyclic free radical (6) obtained by metathesis from the tricyclodecane. In the case of cyclic and polycyclic free radicals, β-scission decompositions lead to new species which carry both a radical center and a double bond. The reverse reaction of addition of the radical center to the double bond can give back the same cyclic radical, but in some cases, also a different one as shown in figure 14 in the case of the cyclohexyl radical. These reactions were also taken in account and written as their reverse decomposition.

FIGURE 14

TABLE 7

The rate constants of β-scission decompositions involving acyclic species or the breaking of a C-H bond are given in Table 6 and are deduced from the values used by EXGAS software (16, 37, 38). The activation energies of Table 6 are not valid for β-scission decompositions leading to the opening of cyclic and polycyclic free radicals. As shown in Tables 8 and 9, the activation energies of β-scission decompositions of cyclohexyl and cyclopentyl radicals are equal to 29.3 and 34.4 kcal.mol$^{-1}$, respectively, whereas the value of Table 6 in the case of the β-scission decomposition of a secondary free radical would be of 27.7 kcal.mol$^{-1}$. That has led us to perform density function theory calculations using CBS-QB3 method (32) with Gaussian03 software (27) to estimate activation energies of 22 model β-scission decompositions involving five- and six-membered ring compounds. As illustrated in Table 7 in the case of the β-scission decompositions of two radicals deriving from metatheses on tricyclodecane, the activation energy of all the β-scission decompositions involving the opening of a cycle have been estimated as being equal to that of one of these model reactions. The values of A factor used in the case of cyclic compounds are the same than the values used by EXGAS for the corresponding reactions; i.e. $2x10^{13}$ $s^{-1}$ for each identical alkylic bond which can be broken and $3.3x10^{13}$ $s^{-1}$ for an allylic bond.

TABLES 8-9

*c/ Propagation by isomerization*

Only isomerizations giving directly a resonance stabilized radical have been considered. As shown in Table 7, no isomerization is then considered for the radical formed by the reaction of metathesis (6) of Figure 13a and only the transfer of the H-atom (x) is taken into account in the case of the deriving radical. As for diradicals, A-factors are mainly based on the changes in the number of internal rotations as the reactant moves to the transition state (17). Activation energies are set equal to the sum of the activation energy for H-abstraction from the substrate by analogous radicals and the strain energy of the cyclic transition state (16, 17). Activation energies for the abstraction of H by •R are taken to equal 6.5 $kcal.mol^{-1}$ (secondary allylic H-atom) and 5.5 $kcal.mol^{-1}$ (tertiary allylic H-atom). Strain energies of the monocyclic transition states are taken to equal 26.0 $kcal.mol^{-1}$ (for a four-membered ring), 6.3 $kcal.mol^{-1}$ (for a five members ring), 1 $kcal.mol^{-1}$ (for a six-membered ring) according to Benson (20). For an isomerization through a polycyclic transition state strain energy used for the estimation of the activation energy of the reaction is the one of the corresponding hydrocarbon (Table 5) (34-36).

## DISCUSSION

Simulations have been performed using the software PSR of CHEMKIN II (20) and the mechanism above described which involves 898 species and includes 2623 reactions. The agreement between computed and experimental mole fractions vs. residence time is globally satisfactory for conversion of tricyclodecane (Figure 2a) and for products such as ethylene, 1,3-cyclopentadiene, cyclopentene, 1,5-hexadiene, toluene and 3-cyclopentyl-cyclopentene (Figures 3 and 4). Computed mole fractions of benzene and hydrogen are slightly under-predicted for the longest residence times. Light products such as methane, ethane and propene are more strongly under-predicted (a factor around 2 for methane and propene, a factor up to 4 for ethane).

Figure 2b displays a comparison between computed and experimental conversions of tricyclodecane as a function of temperature for several initial hydrocarbon mole fractions (0.7, 2 and 4%) and for a residence time of 1 s. Simulations show that computed conversions depend on the initial hydrocarbon mole fraction as it is the case experimentally. Nevertheless computed conversions are over-predicted by a factor lower than 2 for initial hydrocarbon mole fractions of 0.7 and 2%.

Figures 15 and 16 present flow rate analysis performed at 933 K and at a residence time of 1 s corresponding to a conversion of 5.25%. The arrows and the percentages represent the flow of consumption of a species through a given channel. Flow rate analysis shows that 88.4% of the reactant is consumed through metathesis reactions (paths 1-4 in figure 15 and 6-7 in figure 16) with small radicals (hydrogen, methyl and allyl radicals). It underlines also the importance of reactions of unimolecular initiation (path 5 in figure 16) in the consumption of tricyclodecane (11.6%). Unimolecular initiation step leads to the formation of one diradical (path 5) which is quasi-exclusively consumed (99.7%) by a reaction of isomerization (Figure 12c) to form 3-cyclopentyl-cyclopentene. This last species is then consumed by reaction of unimolecular initiation (95.6%) to form a new diradical species which leads to $C_5$ and allyl radicals through successive β-scission decompositions.

The fact that tricyclodecane mainly leads to 3-cyclopentyl-cyclopentene through diradical BR1 (path 1 in Fig. 7) may be surprising because the activation energy of this reaction of unimolecular initiation (77 $kcal.mol^{-1}$) is 6 $kcal.mol^{-1}$ higher than the activation energy of the reaction of unimolecular initiation (2) of Figure 6 as shown in Table 2. If quasi stationary state is assumed for diradicals, the global kinetic constant ($k_g$) of the reaction of formation of a product generated from a diradical can be expressed by the following equation:

$$k_g=(k_1\times k_2)/(k_{-1}+k_2) \qquad \{8\}$$

with : $k_1$, $k_{-1}$ and $k_2$ the rate constants of the unimolecular initiation leading to the given diradical, of the reverse reaction of recombination and of the reaction of β-scission decomposition or of isomerization consuming the diradical, respectively (Figure 9a).

Diradical BR1 reacts mainly by isomerization to form 3-cyclopentyl-cyclopentene. At 1000 K, $k_1$ is equal to $9.5\times10^{-2}$ $s^{-1}$ according to Table 2, $k_{-1}$ is equal to $1.3\times10^{9}$ $s^{-1}$ according to Table 2 and $k_2$ is equal to $3.8\times10^{11}$ $s^{-1}$ according to Table 3. This leads to a global rate constant $k_{g(BR1)}$ of about $9.5\times10^{-2}$ $s^{-1}$. Diradical BR2 is less stable than diradical BR1 as shown in Table 1 and the activation energy of its main isomerization is much larger. At 1000 K, the values of $k_1$, $k_{-1}$ and $k_2$ are respectively equal to 1.9 $s^{-1}$ (Table2), $3.8\times10^{13}$ $s^{-1}$ (Table 1) and $1.3\times10^{9}$ $s^{-1}$ ($A_2 = 9.7\times10^{9}xT$ $s^{-1}$ and $Ea_2 = 17.5$ kcal.mol$^{-1}$), leading to a global rate constant $k_{g(BR2)}$ of $6.5\times10^{-5}$ $s^{-1}$ which is lower than the previous value of a factor 1500.

FIGURES 15-16

As explained in the description of the mechanism six radicals are obtained through reactions of metathesis on the reactant. Radicals obtained through path 1 are mainly consumed to form ethylene and a $C_8$ radical species which leads to the formation of toluene and methyl radicals (45.6%) and to the formation of benzene and hydrogen radical (45.6%). Radicals obtained through path 2 lead to cyclopentadiene and methyl radical (16.0%), cyclopenten-3-yl radical (precursor of cyclopentadiene) and 1,4-pentadiene (8.7%) and 3-ethenyl-cyclopentene through two ways (66.5 and 8.7%). The formation of 3-ethenyl-cyclopentene has not been detected experimentally. This species should react through secondary reactions to produce toluene. Radicals of path 3 are mainly consumed through β-scission decompositions to form ethylene and a new $C_8$ radical that leads to the formation of cyclopentadiene and allyl radical through several isomerizations and β-scission decompositions. Radicals of path 4 lead mainly to cyclopentene and cyclopenten-4-yl radical (precursor of cyclopentadiene). Discussion on path 5 (unimolecular initiation) has been previously presented. Radicals formed through path 6 mainly produce allyl radicals and norbornene. Norbornene reacts easily through retro Diels-Alder reaction (concerted mechanism) to form ethylene and cyclopentadiene (39).

Radicals obtained through path 7 are consumed through two ways, 62.3 % of them give cyclopentadiene and cyclopenten-3-yl and 37.4% ethylene and benzene through several isomerizations and β-scission decompositions.

As there are too many ways of formation of light products such as hydrogen, propene (mainly formed through metatheses of hydrogen radical or of allyl radical on the reactant) and ethylene, Table 10 displays only the main ways of production of cyclopentadiene, cyclopentene, benzene and toluene. Cyclopentadiene comes from path 3 of Figure 15 (28.6%) and from the concerted decomposition of norbornene formed through paths 2 and 6 (28.1%). It is also formed from cyclopenten-4-yl radical (21.6%), cyclopenten-3-yl radical (4.6%) and 3-methyl-cyclopenten-5-yl radical (3.9%). Another source of cyclopentadiene is the molecular decomposition of cyclopentene (10.8%). Cyclopentene is mainly obtained from path 4 (44.4%). It is also formed through path 7 (17.9%) and through the metathesis of the cyclopenten-3-yl radical on the reactant (31.2%). Benzene is mainly obtained though the paths 1 and 7 (respectively 65.6 and 25.2 %). Another source of formation is the reaction of allyl radicals with propargyl radicals (8.3%). The formation of toluene mainly occurs through path 2 (98.9%). Path 5 (unimolecular initiation step) is the only source of formation of 3-cyclopentyl-cyclopentene and 1,5-hexadiene is obtained from the combination of two allyl radicals.

TABLE 10

Figure 17 displays sensitivity analyses related to the mole fractions of tricyclodecane, 3-cyclopentyl-cyclopentene, cyclopentadiene and benzene. They were performed at a temperature of 933 K and at a residence time of 1 s. Sensitivity analysis on tricyclodecane shows that the conversion of the reactant is controlled by three types of reactions: the reaction of unimolecular initiation of tricyclodecane leading to the formation of diradical BR1, the reactions of metathesis of allyl radicals on tricyclodecane and the reaction of unimolecular initiation of 3-cyclopentyl-cyclopentene which leads to the formation of the diradical BR60. The formation of 3-cyclopentyl-cyclopentene is mainly ruled by the reaction of unimolecular initiation of tricyclodecane to BR1. Two reactions have an influence on its consumption: the unimolecular initiation leading to diradical BR60 and to a lesser extent the unimolecular initiation producing radicals R175 and R69. Like the conversion of tricyclodecane, the formation of products, such as cyclopentadiene and benzene, is mainly ruled by the unimolecular initiations of tricyclodecane and 3-cyclopentyl-cyclopentene giving diradicals BR1 and BR60, respectively, and by reactions of metathesis of hydrogen, methyl and allyl radicals on tricyclodecane. The formation of benzene is also influenced by reactions (1.5), (1.6), (6.1) and (6.2) of figures 15-16. Indeed reactions (1.5) and (1.6) are competitive: reaction (1.5) leads to the formation of toluene and methyl radical whereas reaction (1.6) leads to the formation of benzene, ethylene and hydrogen radical. This is the same explanation for reaction (7.1) and (7.2): reaction (7.2) is a way of formation of benzene whereas path (7.1) produces $C_5$ species.

FIGURE 17

According to flow rate and sensitivity analyses some simplifications of the mechanism can be done. As previously explained reactions of unimolecular initiation of the tricyclodecane exclusively lead to the formation of 3-cyclopentyl-cyclopentene and it is possible to consider the only reactions of unimolecular initiation and of isomerization which form this species. It was shown that 3-cyclopentyl-cyclopentene mainly reacts by unimolecular reactions. Reactions of metathesis of radicals on this species (bimolecular reactions) can be neglected because in the conditions of the study presented here its concentration remains low (the rate of consumption of this species is of the same magnitude than its rate of formation). On the contrary reactions of metathesis of radicals on the initial reactant appeared to be very important and can't be simplified. It seems to be difficult to simplify the part of the mechanism related to the reactions of propagation of the radicals obtained from metatheses on the reactant (β-scission and isomerization) because kinetic parameters of these two kinds of reactions are of the same magnitude.

The conversion of tricyclodecane is greatly influenced by unimolecular initiations, that was also true in the case of the thermal decomposition of n-dodecane (18). While an unimolecular initiation of a linear alkane leads directly to the formation two free radicals which can react by metathesis with the reactant and promote chain reactions, the unimolecular initiation step of tricyclodecane leads to a stable molecule, which only at its turn decomposes to give two active radicals. Therefore, despite the fact that unimolecular initiations are easier for strained polycyclic compounds than in linear hydrocarbons, the reactivity of tricyclodecane is much lower than that of n-dodecane as shown in Figure 2c.

The under-estimation observed for light products, such as hydrogen, methane, ethane and propene, is mainly due to the fact that secondary reactions of middle weight primary products have not been comprehensively taken in account at this step of the study. Light hydrocarbons are primarily obtained from the reactant through numerous channels (see fig. 15 and 16), but also from intermediate middle weight products through secondary reactions which are not included in the model presented in this paper (excepted for the primary products which are taken in account in the $C_0$-$C_6$ reaction base). Future work will involve the writing of an n-airy mechanism in which the reactions of primary products (such as 3-ethenyl-cyclopentene for example, whose formation is predicted by the model, but which is not detected experimentally) will be detailed.

## CONCLUSION

The thermal decomposition of tricyclodecane has been studied in a jet-stirred reactor over a range of temperature from 848 K to 933 K and for residence times between 0.5 and 6s. Conversion of the initial hydrocarbon varied between 0.01 and 25%. Eleven products of the reaction were identified. Major products are hydrogen, ethylene, propene and cyclopentadiene. Products such as cyclopentadiene, benzene and toluene which are often secondary products seem to have here primary sources.

A comprehensive detailed kinetic model has been written using a systematic method and by following a diradical approach for the unimolecular initiation step of tricyclodecane. Kinetic parameters were taken from literature when available or deduced from estimation performed by theoretical density function theory calculations in the case of reactions involving cyclic compounds and diradicals. Agreement between experimental and computed mole fractions is satisfactorily for both the conversion of the reactant and most products of the reaction. The under-estimation observed for some light products underlines the need to write a n-airy mechanism in which the reactions of primary products will be more comprehensively detailed.

Flow rate and sensitivity analyses were performed at a temperature of 633 K and at a residence time of 1 s. The results of these analyses underlined the importance of the reactions of unimolecular initiation of tricyclodecane and of 3-cyclopentyl-cyclopentene for conversion of the initial hydrocarbon. Another future work would be a reduction of this comprehensive, but certainly in some ways uselessly detailed mechanism, by using the results of flow rate and sensitivity analyses.

ACKNOWLEDGMENTS

This work was supported by MBDA-France and the CNRS. The authors are very grateful to E. Daniau, M. Bouchez and F. Falempin for helpful discussions.

**TABLE 1** : Thermodynamic data of the 7 diradicals directly obtained from tricyclodecane (see Figure 6) and for the 2 radicals and the 5 diradicals deriving from 3-cyclopentyl-cyclopentene (see Figure 7). Values in bold have been determined by quantum calculations. (units: $kcal.mol^{-1}$ for ΔHf°, $cal.mol^{-1}.K^{-1}$ for S° and Cp°).

| *Species* | *ΔHf°* (298 K) | *S°* (298 K) | *Cp°* (300 K) | (400 K) | (500 K) | (600 K) | (800 K) | (1000 K) | (1500 K) |
|---|---|---|---|---|---|---|---|---|---|
| Tricyclo-decane | -14.38 | 85.19 | 37.46 | 54.24 | 68.28 | 80.22 | 97.89 | 109.77 | 125.23 |
| BR1 | 57.41 | 106.36 | 39.60 | 55.17 | 68.25 | 79.29 | 96.23 | 107.93 | 126.63 |
| | **58.3** | **98.4** | **42.7** | **57.5** | **70.7** | **81.6** | **98.1** | **109.8** | **127.5** |
| BR2 | 61.79 | 98.37 | 40.88 | 57.71 | 71.78 | 83.99 | 102.88 | 115.56 | 134.73 |
| BR3 | 67.49 | 106.03 | 39.92 | 55.95 | 69.87 | 81.39 | 99.04 | 111.64 | 131.36 |
| BR4 | 67.39 | 103.59 | 40.14 | 56.34 | 70.25 | 81.60 | 99.14 | 111.60 | 131.31 |
| BR5 | 67.80 | 106.61 | 44.96 | 60.61 | 73.85 | 84.79 | 101.37 | 113.10 | 127.51 |
| BR6 | 72.21 | 105.65 | 41.65 | 56.72 | 69.57 | 79.96 | 96.32 | 107.79 | 124.91 |
| BR7 | 68.14 | 105.56 | 45.17 | 60.99 | 74.23 | 85 | 101.47 | 113.06 | 127.46 |
| MA110[a] | -0.74 | 102.69 | 40.02 | 55.40 | 68.62 | 79.87 | 97.35 | 109.55 | 128.26 |
| R69 | 38.55 | 70.36 | 17.68 | 24.25 | 29.94 | 34.87 | 42.74 | 48.37 | 56.54 |
| R175 | 23.94 | 74.41 | 19.57 | 27.69 | 34.61 | 40.53 | 49.82 | 56.40 | 66.26 |
| BR60 | 51.49 | 117.81 | 51.94 | 69.11 | 83.88 | 96.51 | 116.28 | 130.23 | 149.98 |
| BR61 | 63.72 | 116.96 | 43.13 | 57.75 | 70.26 | 80.92 | 97.63 | 109.59 | 128.44 |
| BR62 | 78.38 | 121.55 | 47.79 | 61.93 | 73.94 | 84.16 | 100.29 | 111.98 | 129.62 |
| BR63 | 78.82 | 119.49 | 47.80 | 62.12 | 74.34 | 84.74 | 101.04 | 112.72 | 130.36 |
| BR64 | 80.06 | 119.49 | 47.80 | 62.12 | 74.34 | 84.74 | 101.04 | 112.72 | 130.36 |

[a] MA110 corresponds to the molecule of 3-cyclopentyl-cyclopentene.

**TABLE 2** : Activation energies of the unimolecular initiations by breaking of C-C bonds from tricyclodecane (ring strain energy of 22 $kcal.mol^{-1}$) and 3-cyclopentyl-cyclopentene (ring strain energy of 12 $kcal.mol^{-1}$). The reactions in bold are those with the higher flow rate under the conditions of our study.

| Unimolecular initiation (see figures 6 and 7) | Diradical ring strain energy ($kcal.mol^{-1}$) | Lost ring strain energy ($kcal.mol^{-1}$) | Activation energy ($kcal.mol^{-1}$) |
|---|---|---|---|
| from tricyclodecane (see figure 6) | | | |
| **1** | **12[a]** | **10** | **77** |
| 2 | 6[b] | 16 | 71 |
| 3 | 12[a] | 10 | 77 |
| 4 | 12[a] | 10 | 77 |
| 5 | 16[c] | 6 | 81 |
| 6 | 16[c] | 6 | 81 |
| 7 | 16[c] | 6 | 81 |
| from 3-cyclopentyl-cyclopentene (see figure 7) | | | |
| (1)[d] | - | - | 71.2[f] |
| **2** | **6[e]** | **6** | **65.2[f]** |
| 3 | 6[e] | 6 | 65.2[f] |
| 4 | 6[e] | 6 | 81 |
| 5 | 6[e] | 6 | 81 |
| 6 | 6[e] | 6 | 81 |

[a] Ring strain energy taken equal to twice that of cyclopentane according to Benson (20).

[b] Ring strain energy taken equal to the sum of those of cyclopentane and cyclohexane according to Benson (20).

[c] Ring strain energy taken equal to that of norbornane according to Benson (20).

[d] Formation of two cyclic radical species.

[e] Ring strain energy taken equal to the one of cyclopentane or of the one of cyclopentene according to Benson (20).

[f] Activation energy for a broken bond in position β of a $Csp^2$-$Csp^2$ bond taken equal to that of the decomposition of 1-pentene to give allyl and ethyl radicals (8).

**TABLE 3** : Products and rate constants of the β-scissions and isomerizations written for the diradical formed by the initiation (1) of Figure 6. Activation energies are given in $kcal.mol^{-1}$ and A-factors in $s^{-1}$.

| Product | Reaction pathway | A | b | Ea |
|---|---|---|---|---|
| Diradical obtained from tricyclodecane 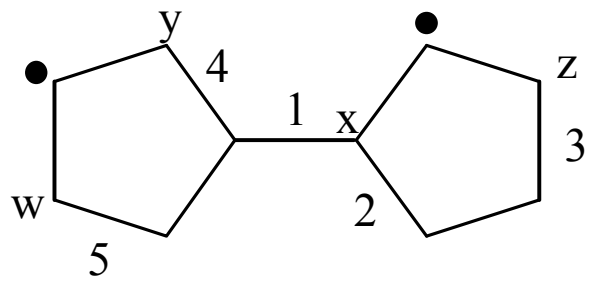 | | | | |
| 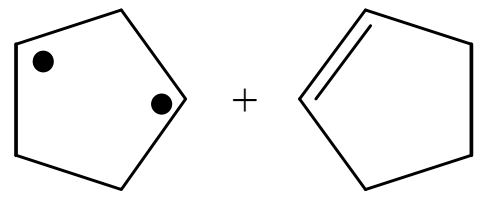 | Breaking of bond 1 | $2.0\times10^{13}$ | 0 | 30.4[a] |
| 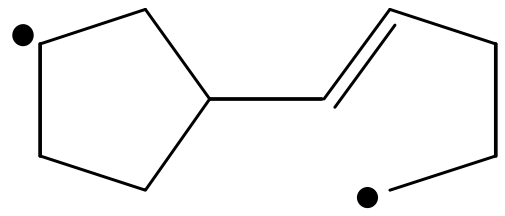 | Breaking of bond 2 | $2.0\times10^{13}$ | 0 | 35.9[b] |
| 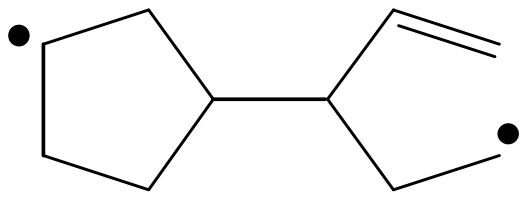 | Breaking of bond 3 | $2.0\times10^{13}$ | 0 | 37.5[c] |
| 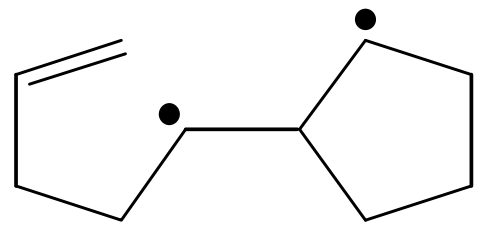 | Breaking of bond 4 | $2.0\times10^{13}$ | 0 | 34.8[d] |
| 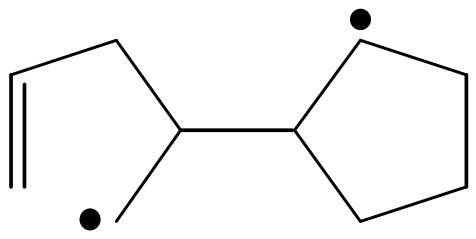 | Breaking of bond 5 | $2.0\times10^{13}$ | 0 | 37.7[e] |
| 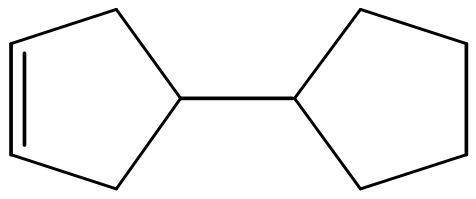 | Isomerization by transfer of H-atom w through a 6 members ring for the transition state | $1.9\times10^{10}$ | 1 | 19.85 |
| 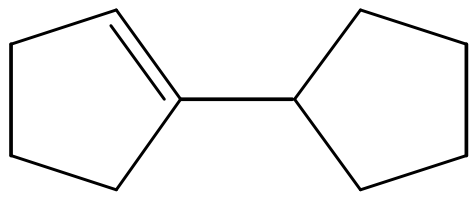 | Isomerization by transfer of H-atom x through a 5 members ring for the transition state | $9.7\times10^{9}$ | 1 | 17.75 |
| 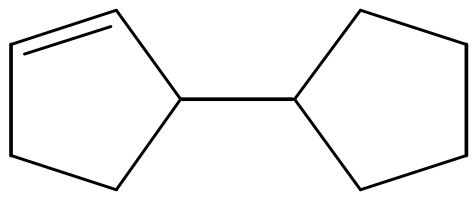 | Isomerization by transfer of H-atoms y or z through a 5 or a 7 members ring for the transition state | y $1.9\times10^{10}$ | 1 | 7.75 |
| | | z $1.9\times10^{10}$ | 1 | 27.75 |

[a] Activation energy taken equal to that of model reaction 7 in Table 8.

[b] Activation energy taken equal to that of model reaction 4 in Table 8.

[c] Activation energy taken equal to that of model reaction 8 in Table 8.

[d] Activation energy taken equal to that of model reaction 9 in Table 8.

[e] Activation energy taken equal to that of model reaction 10 in Table 8.

**TABLE 4** : Products and rate constants of the β-scissions and isomerizations written for the diradical formed by the initiation (2) of Figure 7. Activation energies are given in kcal.mol$^{-1}$ and A factors in s$^{-1}$.

| Product | Reaction pathway | A | b | Ea |
|---|---|---|---|---|
| Diradical BR60 obtained from 3-cyclopentyl-cyclopentene 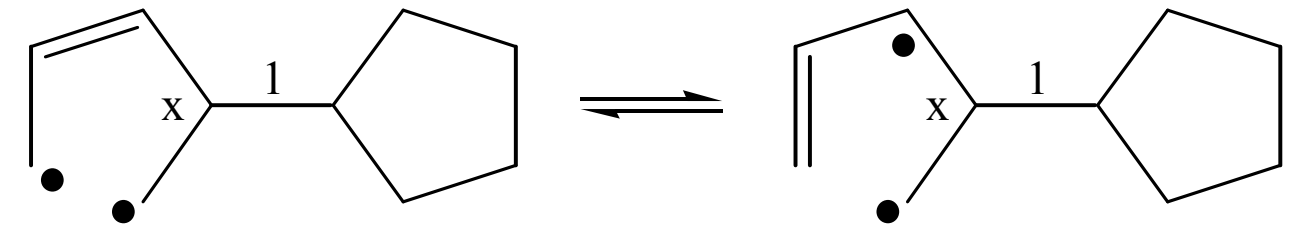 | | | | |
| 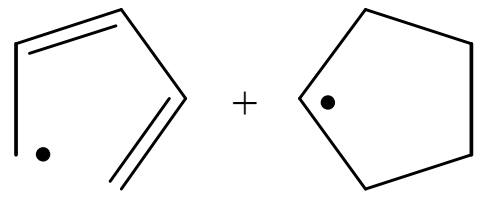 | Breaking of bond 1 | $3.3\times10^{13}$ | 0 | 22.5[a] |
| 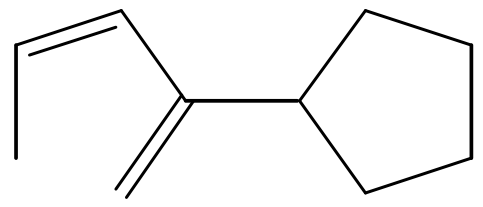 | Isomerization by transfer of H-atom x through a 5 members ring for the transition state | $2.9\times10^{8}$ | 1 | 17.75[b] |
| 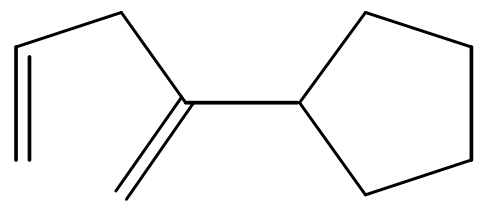 | Isomerization by transfer of H-atom x through a 3 members ring for the transition state | $1.7\times10^{9}$ | 1 | 19.50[b] |
| 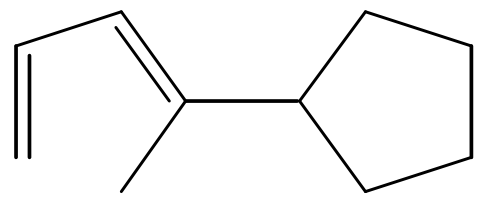 | Isomerization by transfer of H-atom x through a 3 members ring for the transition state | $2.9\times10^{8}$ | 1 | 9.50 |

[a] Breaking of a bond in β position of two radical points as shown in Figure 10b (Ea=22.5 kcal.mol$^{-1}$).

[b] 10 kcal.mol$^{-1}$ were added to the activation energy because of initial resonance stabilized radical.

**Table 5 :** Ring strain energies for some cyclic and polycyclic hydrocarbons.

| Name of the (poly)cyclic species | Ring strain energy ($kcal.mol^{-1}$) |
|---|---|
| cyclobutane | 26[a] |
| cyclopentane | 6[a] |
| cyclohexane | 0[a] |
| bicyclo[2.2.1]heptane (norbornane) | 16[a] |
| bicyclo[3.2.1]octane | 16[b] |
| bicyclo[4.2.1]nonene | 26[b] |
| tricyclodecane | 22[c] |

[a] Ring strain energy according to Benson (20).

[b] Ring strain energy calculated by Maier and Schleyer Von Rague (36).

[c] Ring strain energy calculated by Boyd et al. (31).

**TABLE 6**: Quantitative Structure-Reactivity Relationships. Rate constants are expressed in the form $k = A \times T^{b} \times \exp(-E/RT)$ (with the units $cm^3$, mol, s, kcal) by H atoms which can be abstracted or by bonds which can be broken. $\bullet R_p$, $\bullet R_s$ and $\bullet R_{res.\ stab.}$ are primary, secondary and resonance stabilized free radicals, respectively.

**H-abstraction reactions**

| | Secondary alkylic H | | | Tertiary alkylic H | | | Secondary allylic H | | | Tertiary allylic H | | |
|---|---|---|---|---|---|---|---|---|---|---|---|---|
| | lg A | b | E | lg A | b | E | lg A | b | E | lg A | b | E |
| •H | 6.65 | 2 | 5.0 | 6.62 | 2 | 2.4 | 4.43 | 2.5 | -1.9 | 4.40 | 2.5 | -2.79 |
| $\bullet CH_3$ | 11 | 0 | 9.6 | 11 | 0 | 7.9 | 10.7 | 0 | 7.3 | 10.7 | 0 | 5.6 |
| $\bullet C_2H_5$ | 11 | 0 | 11.0 | 11 | 0 | 9.2 | 0.34 | 3.5 | 4.14 | 0.34 | 3.5 | 2.34 |
| $\bullet R_{res.\ stab.}$ | 1.6 | 3.3 | 18.17 | 1.6 | 3.3 | 17.17 | 1.6 | 3.3 | 18.17 | 1.6 | 3.3 | 17.17 |

**Beta-scission reactions which do not involve a ring**

| | lg A | b | E |
|---|---|---|---|
| Beta-scission of a free radical to | | | |
| $\bullet CH_3$ + alcene | 13.3 | 0 | 31.0 |
| $\bullet R_p$+ alcene | 13.3 | 0 | 28.7 |
| $\bullet R_s$+ alcene | 13.3 | 0 | 27.7 |
| $\bullet R_{res.\ stab.}$+ alcene | 13.5 | 0 | 22.5 |
| $\bullet CH_3$ + diene | 13.1 | 0 | 38.2 |
| •H(secondary) from an alkyl radical | 13.2 | 0 | 34.8 |
| •H(tertiary) from an alkyl radical | 13.2 | 0 | 34.3 |
| •H(primary) from an allylic radical | 13.2 | 0 | 51.5 |
| •H(allylic) from an allylic radical | 13.2 | 0 | 40.7 |
| •H(vinylic) from an allylic radical | 13.1 | 0 | 60.0 |

**TABLE 7** : List of the first reactions of propagation (β-scission decompositions and isomerization) written for the radical formed by the reaction of metathesis (6) of Figure 13 and their rate constants. Activation energies are given in $kcal.mol^{-1}$ and A-factors in $s^{-1}$.

| Product | Reaction pathway | A | b | Ea |
|---|---|---|---|---|
| Radical directly obtained from metatheses on tricyclodecane 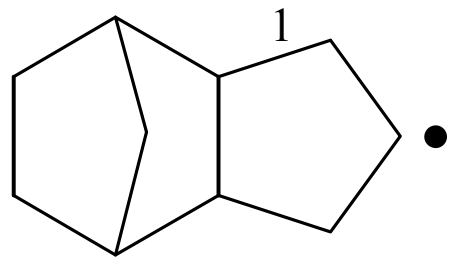 | | | | |
| 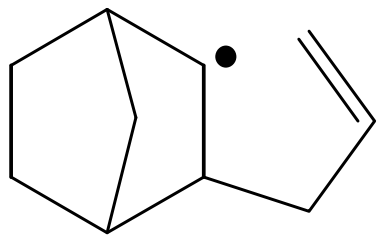 | Breaking of bond 1 | $4.0\times10^{13}$ [a] | 0 | 34.80[b] |
| Radical obtained from β-scission of the previous radical 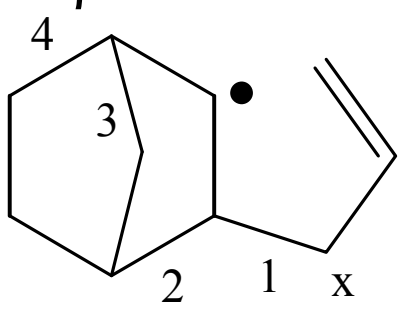 | | | | |
| 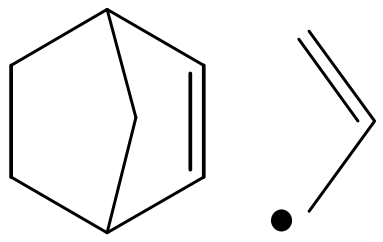 | Breaking of bond 1 | $3.3\times10^{13}$ | 0 | 25.20[c] |
| 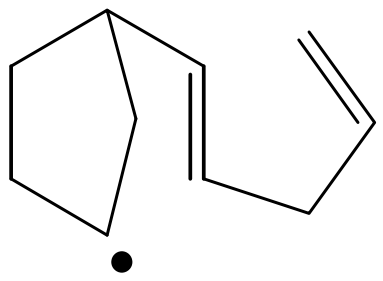 | Breaking of bond 2 | $2.0\times10^{13}$ | 0 | 35.90[d] |
| 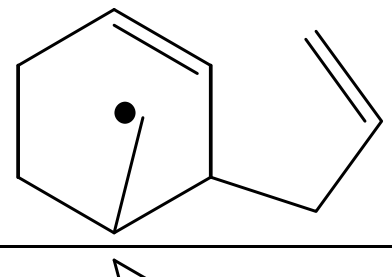 | Breaking of bond 3 | $2.0\times10^{13}$ | 0 | 35.90[d] |
| 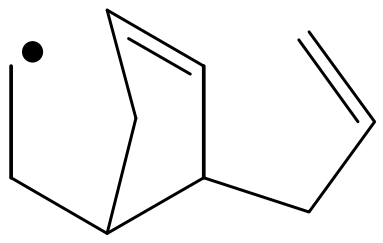 | Breaking of bond 4 | $2.0\times10^{13}$ | 0 | 35.90[d] |
| 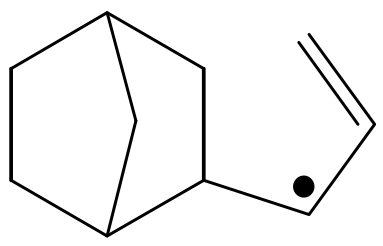 | Isomerization by transfer of H-atom x through a 4 members ring for the transition state | $3.3\times10^{9}$ | 1 | 32.50[e] |

[a] Two identical bonds can be broken .

[b] Activation energy taken equal to that of model reaction 9 in Table 8.

[c] Activation energy taken equal to that of model reaction 6 in Table 8 with 6.2 $kcal.mol^{-1}$ subtracted because of the formation of an allylic radical.

[d] Activation energy taken equal to that of model reaction 4 in Table 8.

[e] Presence of side rings neglected.

**TABLE 8** : Activation energies of model β-scission decompositions by fission of a C-C bond of five-membered ring species calculated by quantum methods. Values in bold are from the literature.

| | Reaction | Activation energy (kcal.mol-1) |
|---|---|---|
| 1 | 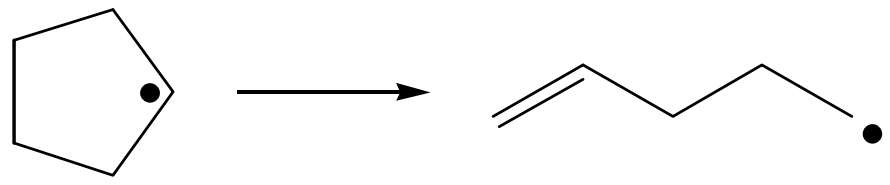 | 33.5<br>**34.4**[a] |
| 2 | 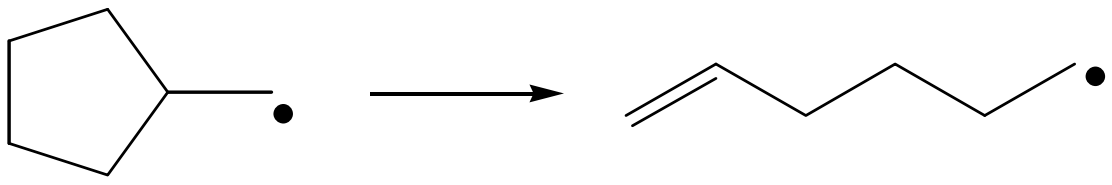 | 24.1 |
| 3 | 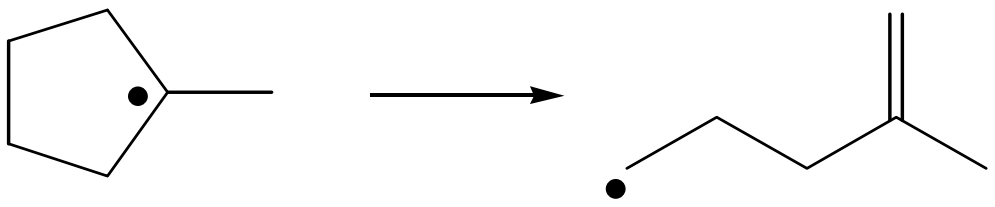 | 37.8 |
| 4 | 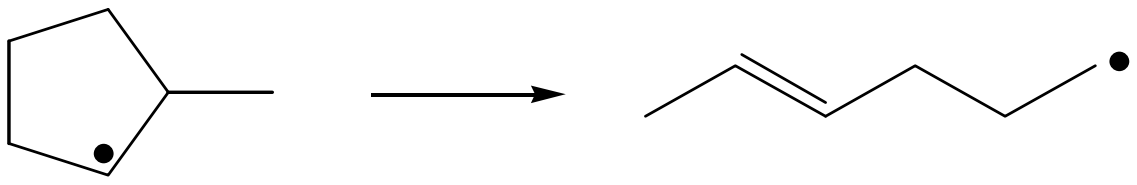 | 35.9 |
| 5 | 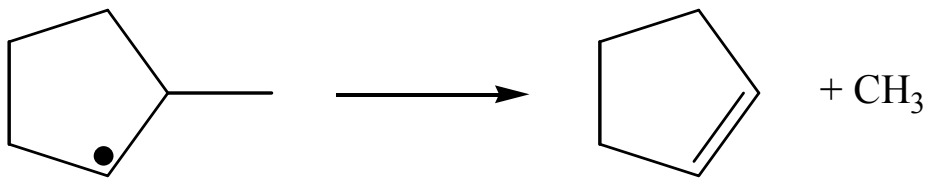 | 33.7 |
| 6 | 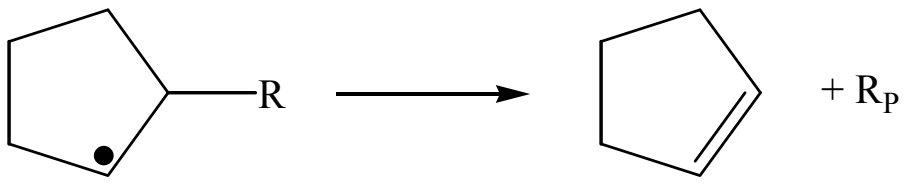 | 31.4[b] |
| 7 | 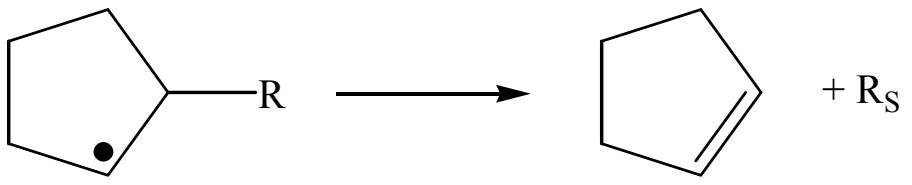 | 30.4[b] |
| 8 | 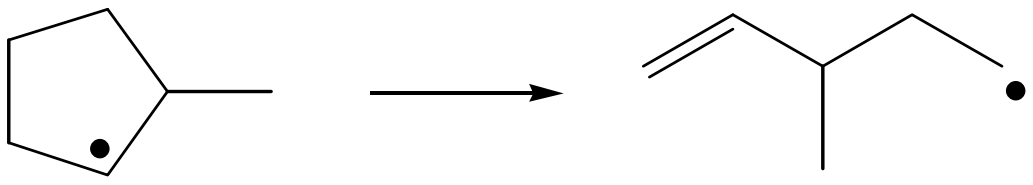 | 37.5 |
| 9 | 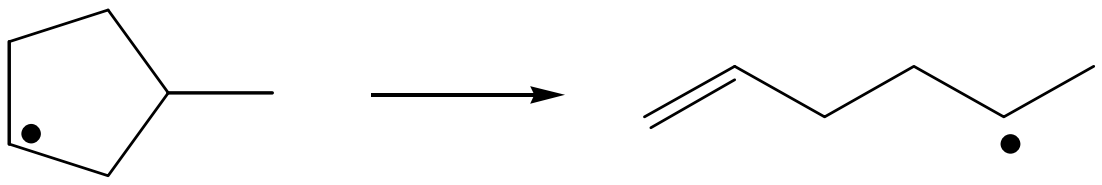 | 34.8 |
| 10 | 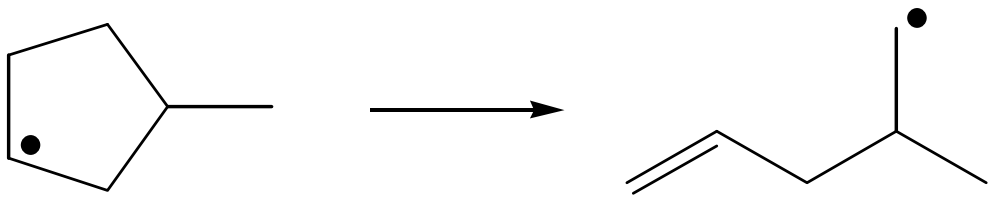 | 37.7 |
| 11 | 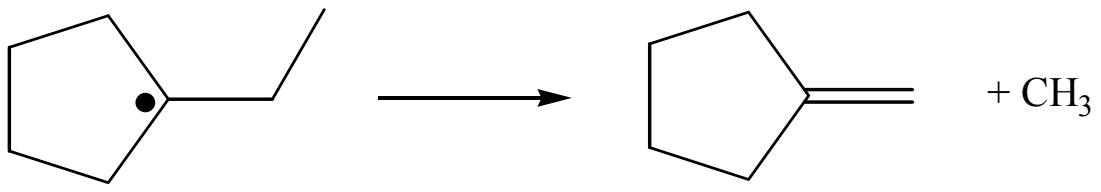 | 34.3 |
| 12 | 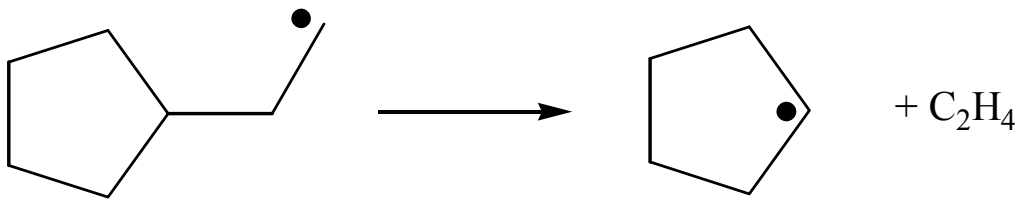 | 26.3<br>**27.7**[c] |

[a] Activation energy from (8)

[b] Activation energy of model reaction 6 and 7 deduced from model reaction 5 and from the correlations used in the Exgas software (16,17).

c Activation energy used in the Exgas software (17)

**TABLE 9** : Activation energies of model β-scission decompositions by fission of a C-C bond of six-membered ring species calculated by quantum methods. Values in bold are from the literature.

| | Reaction | Activation energy ($kcal.mol^{-1}$) |
|---|---|---|
| 1 | 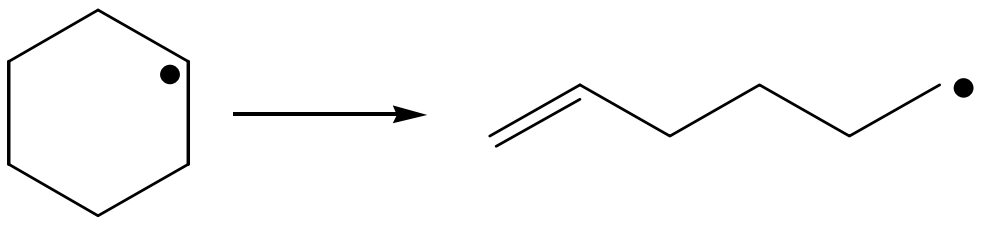 | 31.0<br>**29.3**[a] |
| 2 | 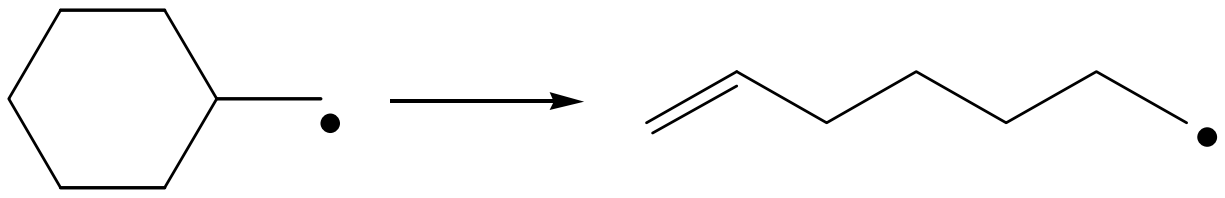 | 19.2 |
| 3 | 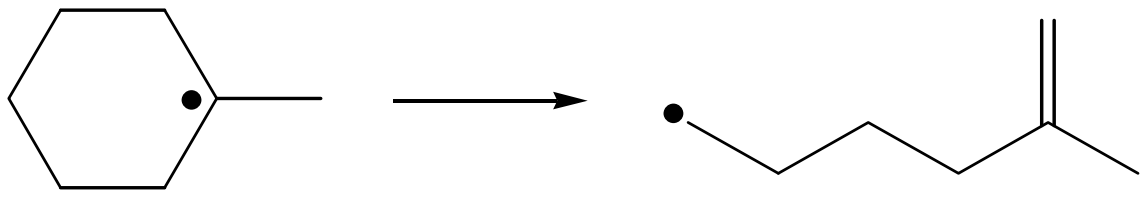 | 35.3 |
| 4 | 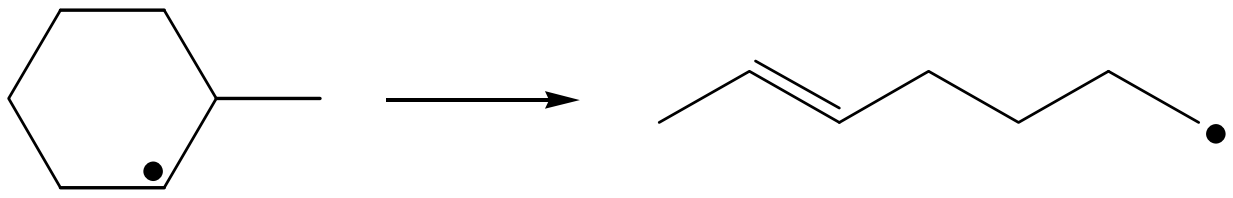 | 33.4 |
| 5 | 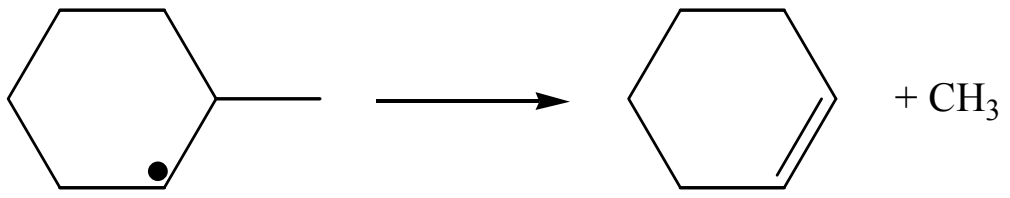 | 31.2 |
| 6 | 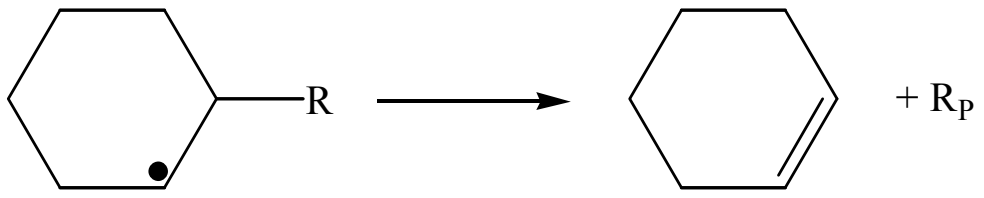 | 28.9[b] |
| 7 | 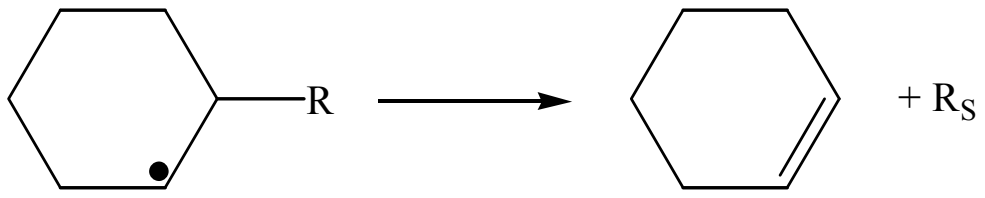 | 27.9[b] |
| 8 | 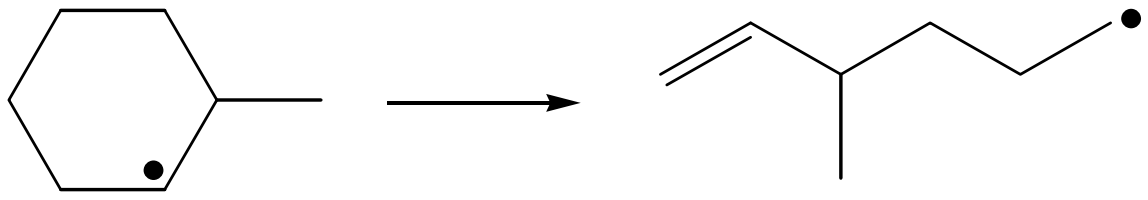 | 35.0 |
| 9 | 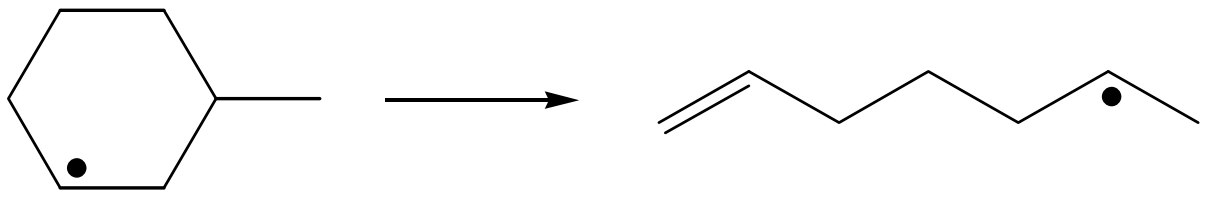 | 32.3 |
| 10 | 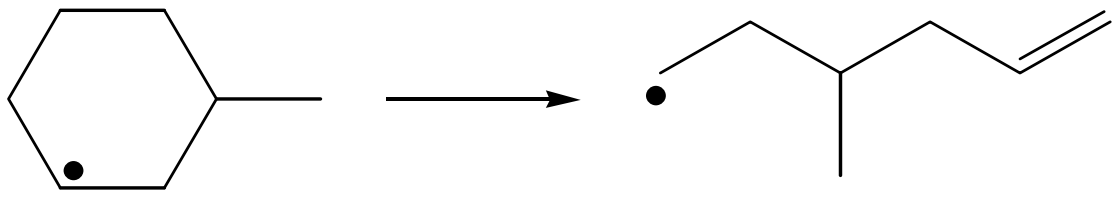 | 35.2 |

[a] Value deduced from the activation energy of the reverse reaction (40).

[b] Activation energy of model reaction 6 and 7 deduced from model reaction 5 and from the correlations used in the Exgas software (16).

**Table 10** : Formation flow analysis for cyclopentadiene, cyclopentene, benzene and toluene (see Figure 15 and Figure 16).

| Species | Channel | % of production through the channel |
|---|---|---|
| $C_5H_6$ | 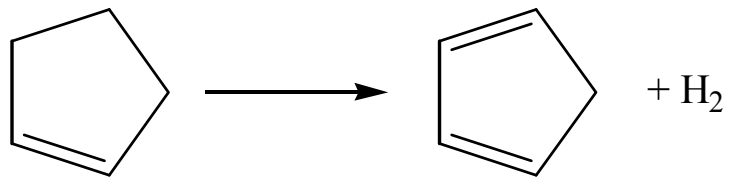 | 10.8 |
| | 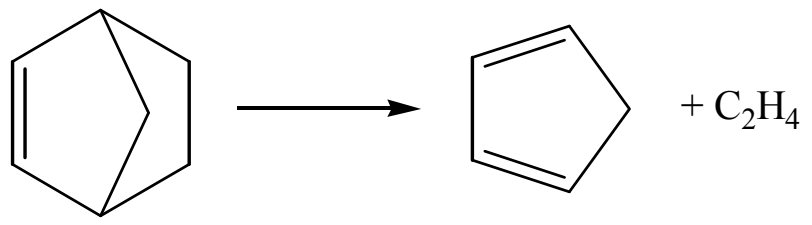 | 28.1 |
| | 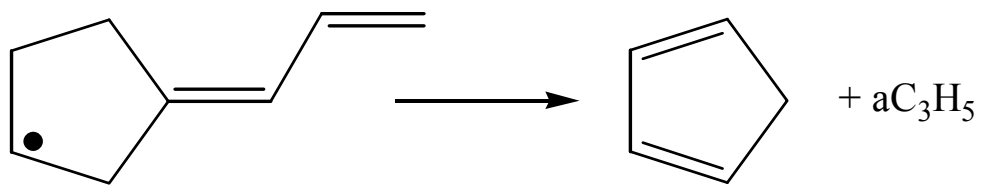 | 28.6 |
| | 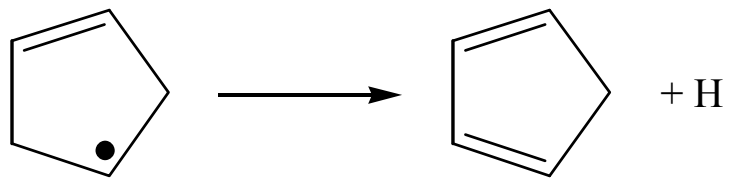 | 21.6 |
| | 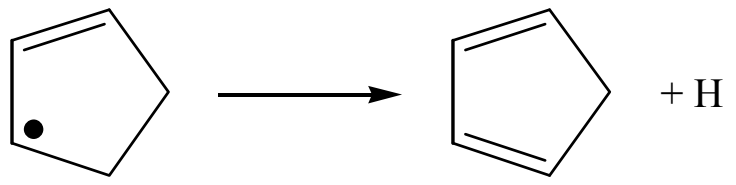 | 4.6 |
| | 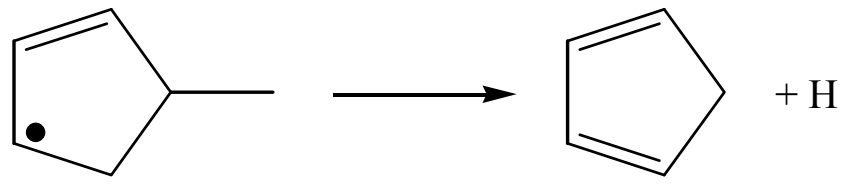 | 3.9 |
| $C_5H_8$ | 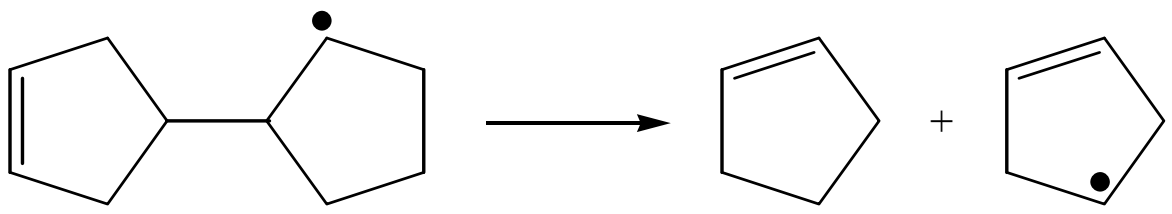 | 44.4 |
| | 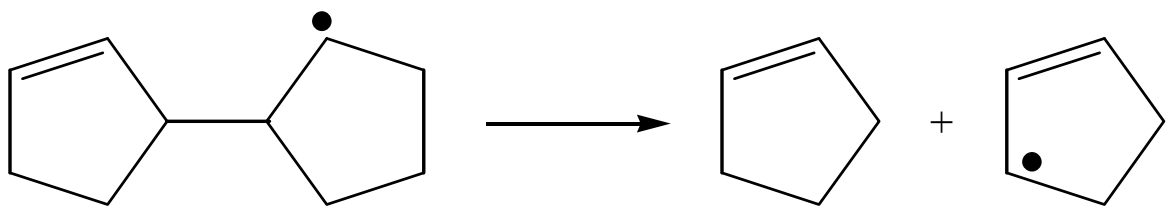 | 17.9 |
| | metatheses of 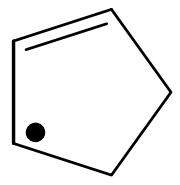 on the reactant | 31.2 |
| | other reactions | 6.0 |
| benzene | 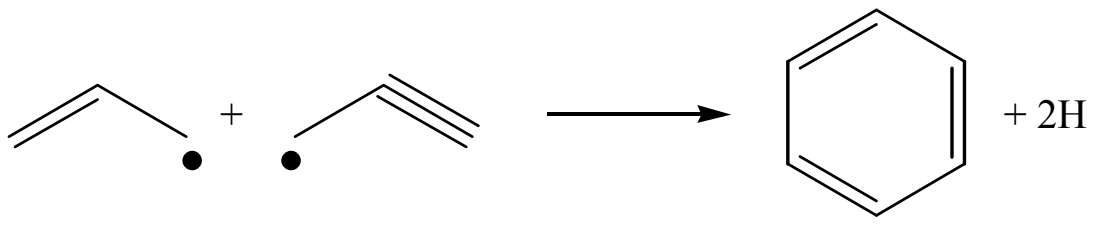 | 8.3 |
| | 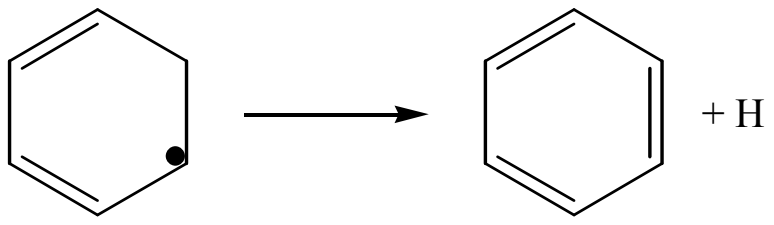 | 90.8 |
| toluene | 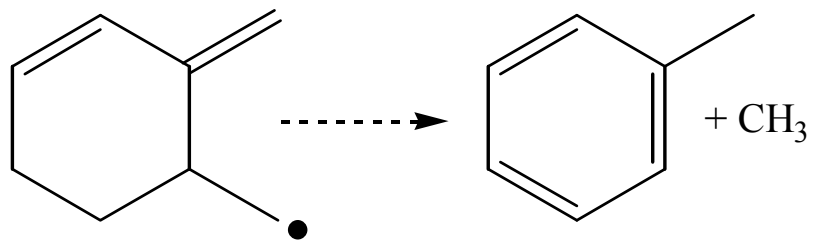 | 98.9 |

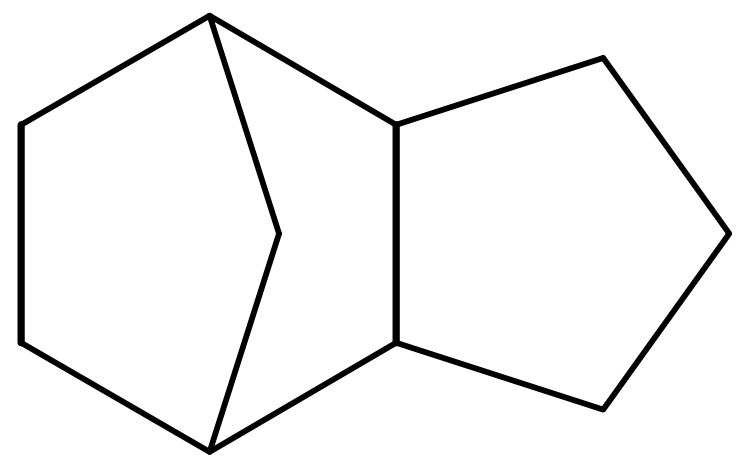

(a)

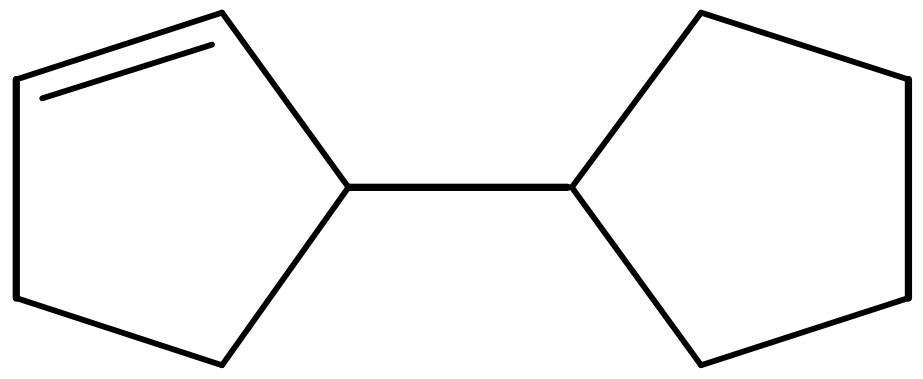

(b)

Figure 1: Structure of (a) exo-tricyclo[5.2.1.$0^{2,6}$]decane, also called exo-tetrahydrodicyclopentadiene or tricyclodecane and of (b) 3-cyclopentyl-cyclopentene.

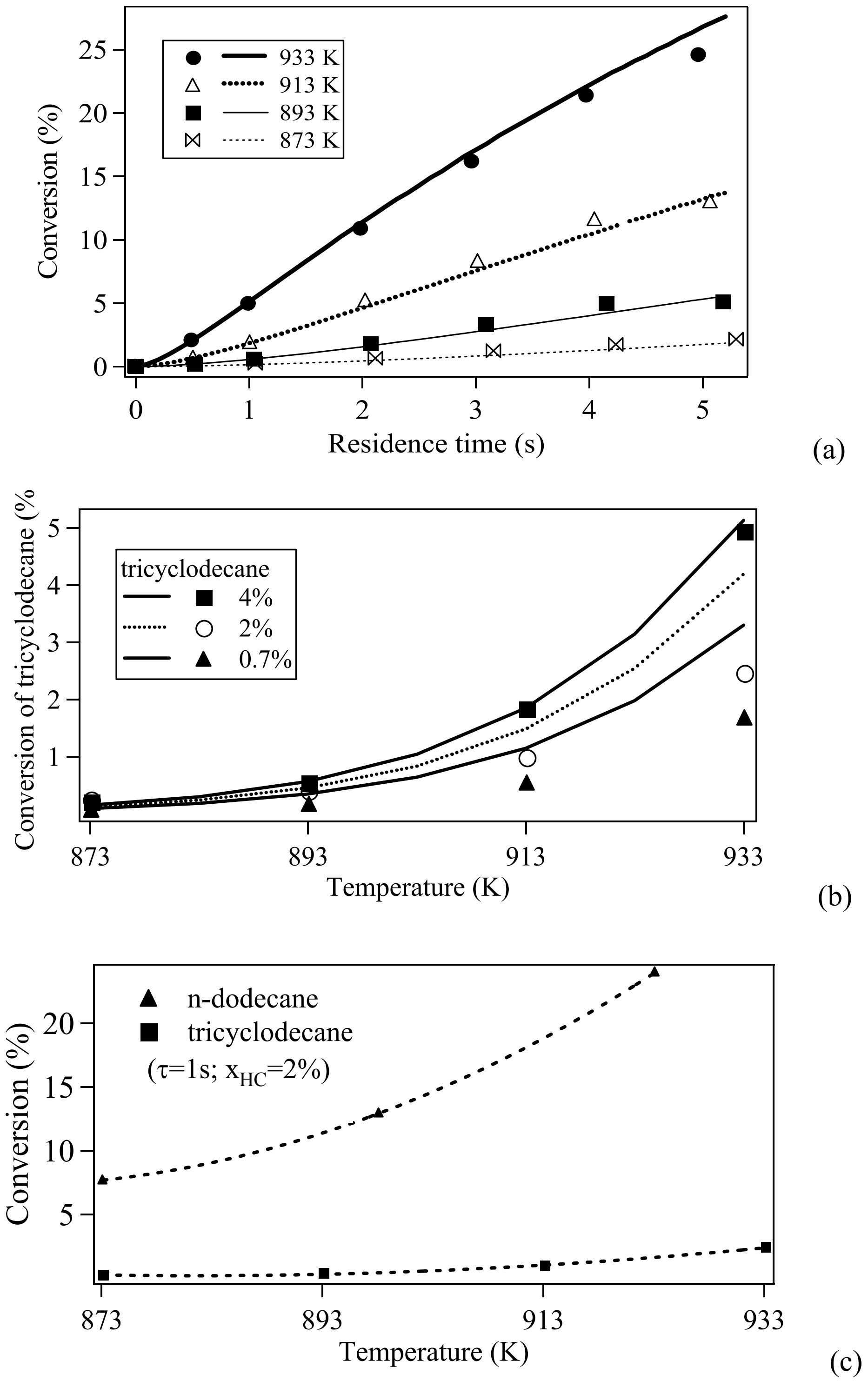

Figure 2: Conversion of tricyclodecane vs. residence time at several temperatures for an initial hydrocarbon mole fraction of 4 % (fig. 2a) and vs. temperature at a residence time of 1s and for several initial hydrocarbon mole fractions (fig. 2b). Comparison with the experimental conversion of n-dodecane (14) vs. temperature at a residence time of 1s and 2 % initial hydrocarbon mole fraction (fig. 2c).

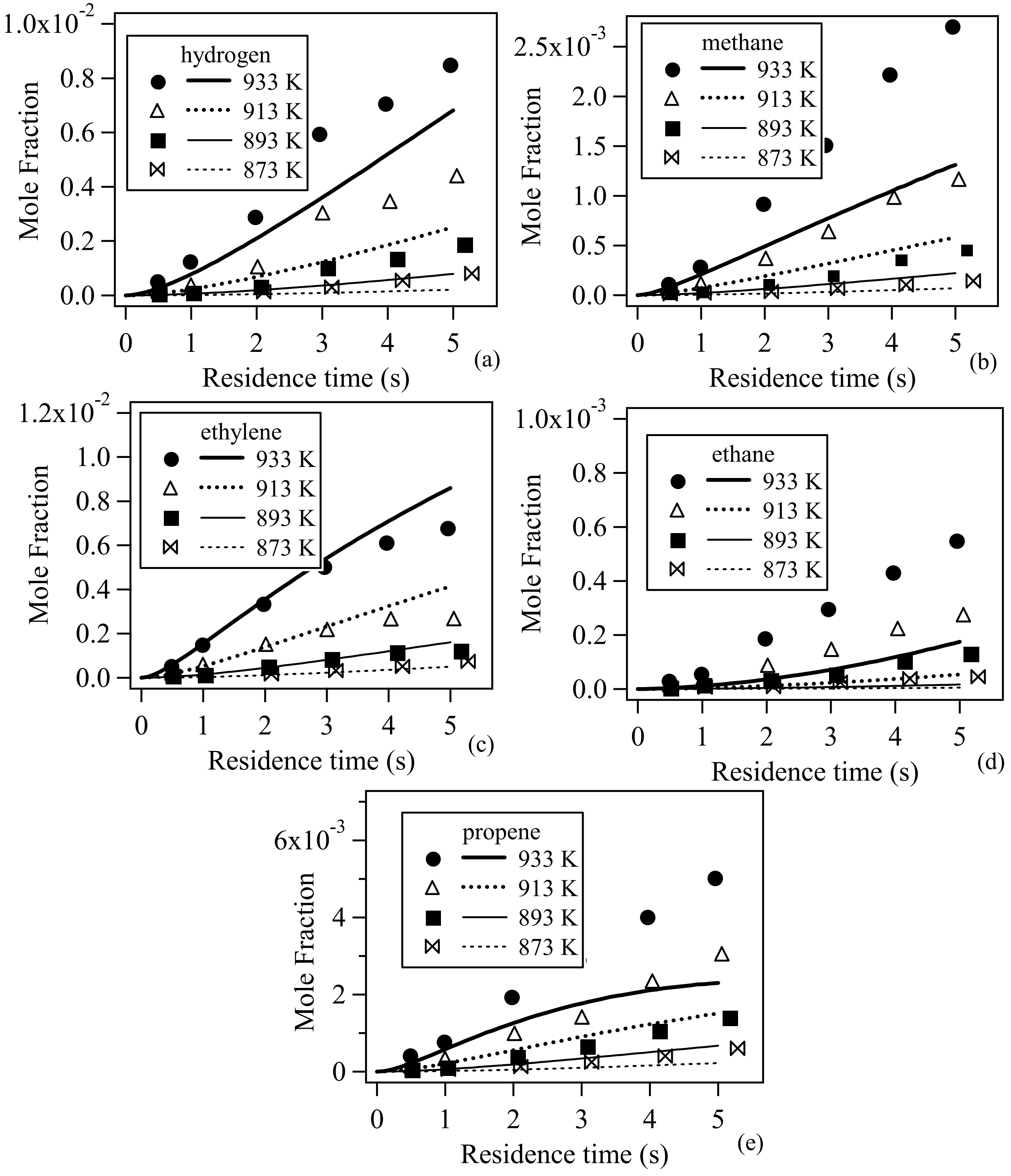

Figure 3: Mole fractions of light products vs. residence time at several temperatures for an initial hydrocarbon mole fraction of 4 %.

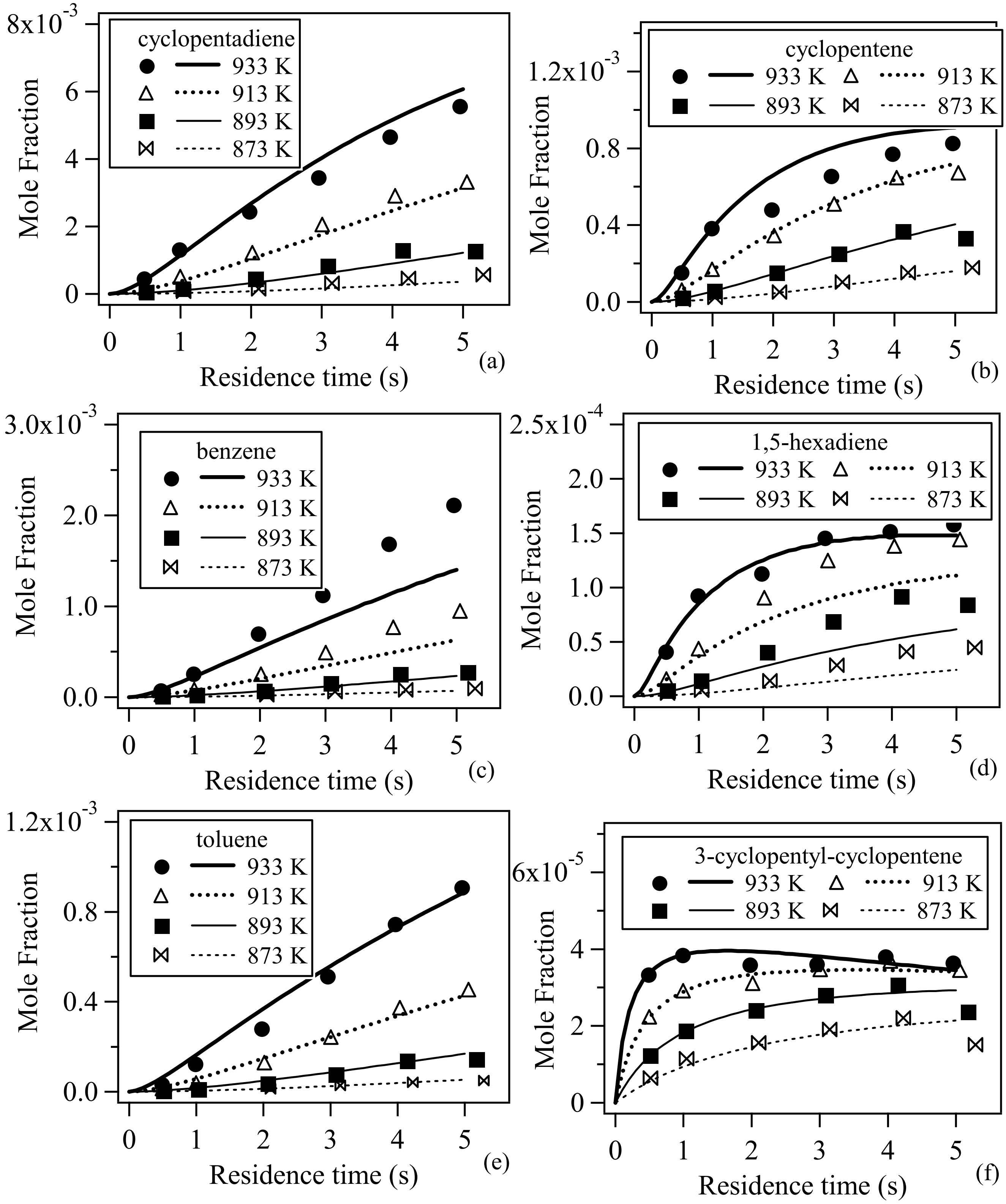

Figure 4: Mole fractions of heavy products vs. residence time at several temperatures for an initial hydrocarbon mole fraction of 4 %.

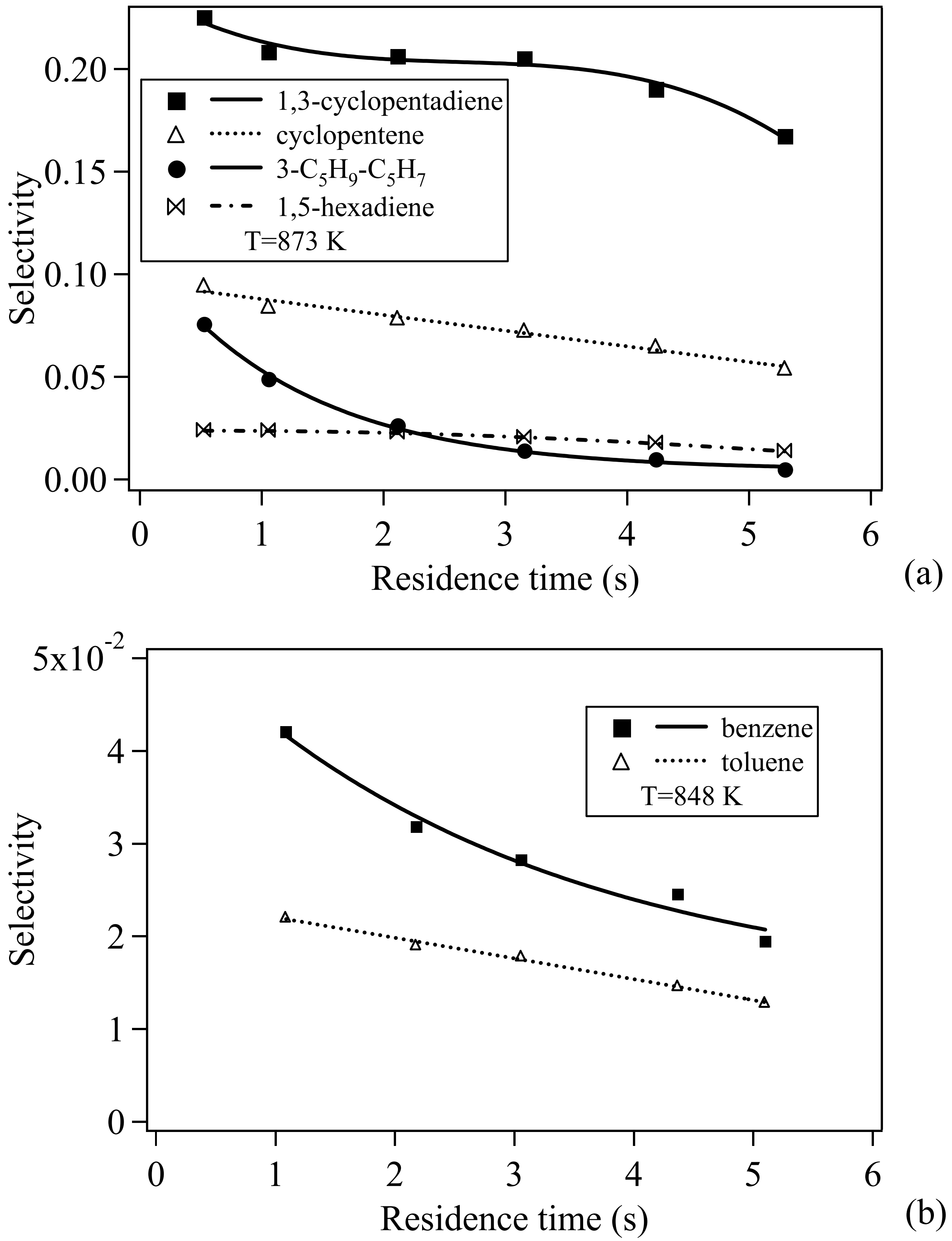

Figure 5: Selectivity of heavy products vs. residence time: (a) 1.3-cyclopentadiene, cyclopentene, 1,5-hexadiene and 3-cyclopentyl-cyclopentene at 873 K and (b) benzene and toluene at 848 K (initial hydrocarbon mole fraction of 4 %).

.

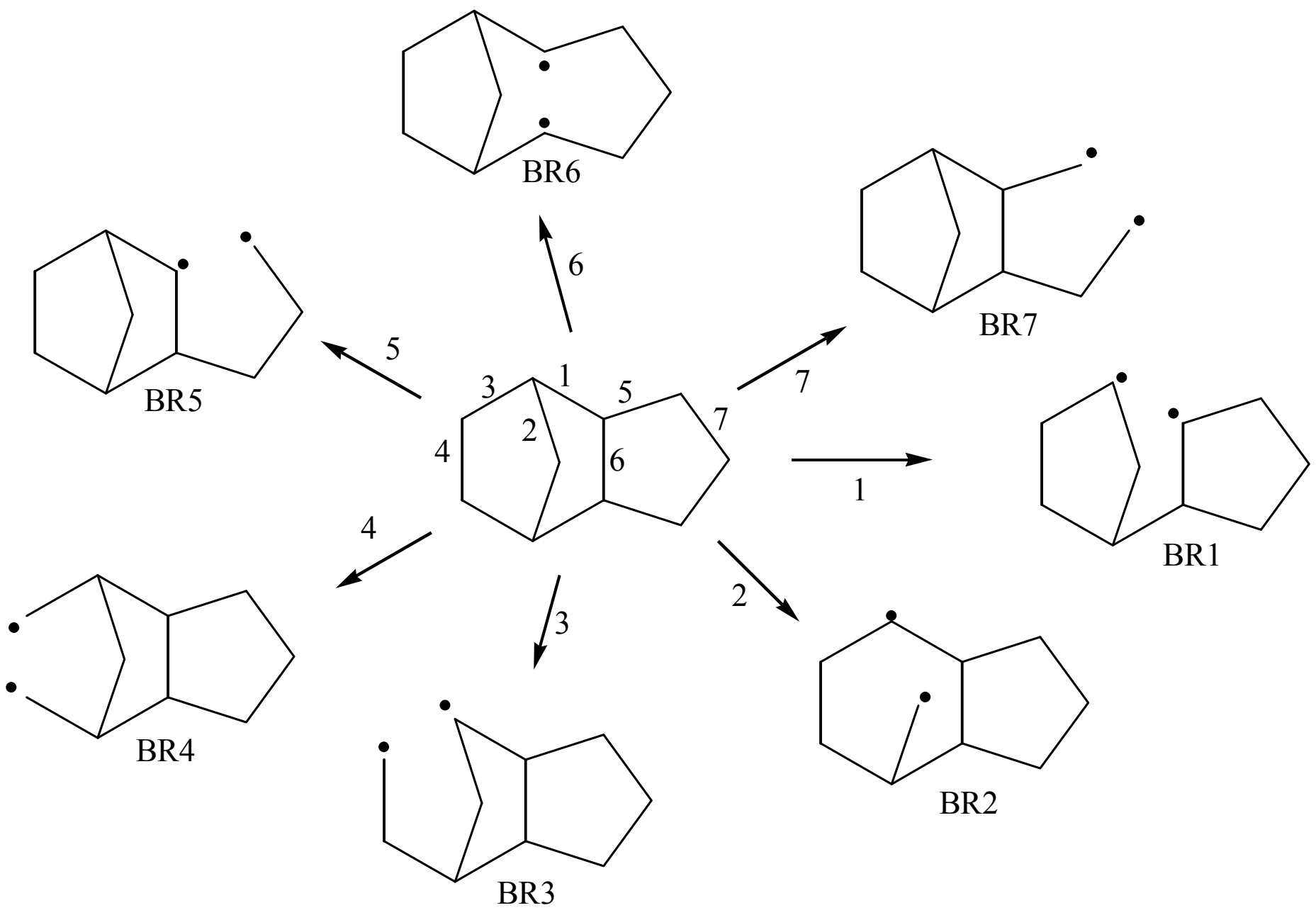

Figure 6: The 7 diradicals which can be obtained from tricyclodecane by unimolecular decomposition involving the breaking of a C-C bond.

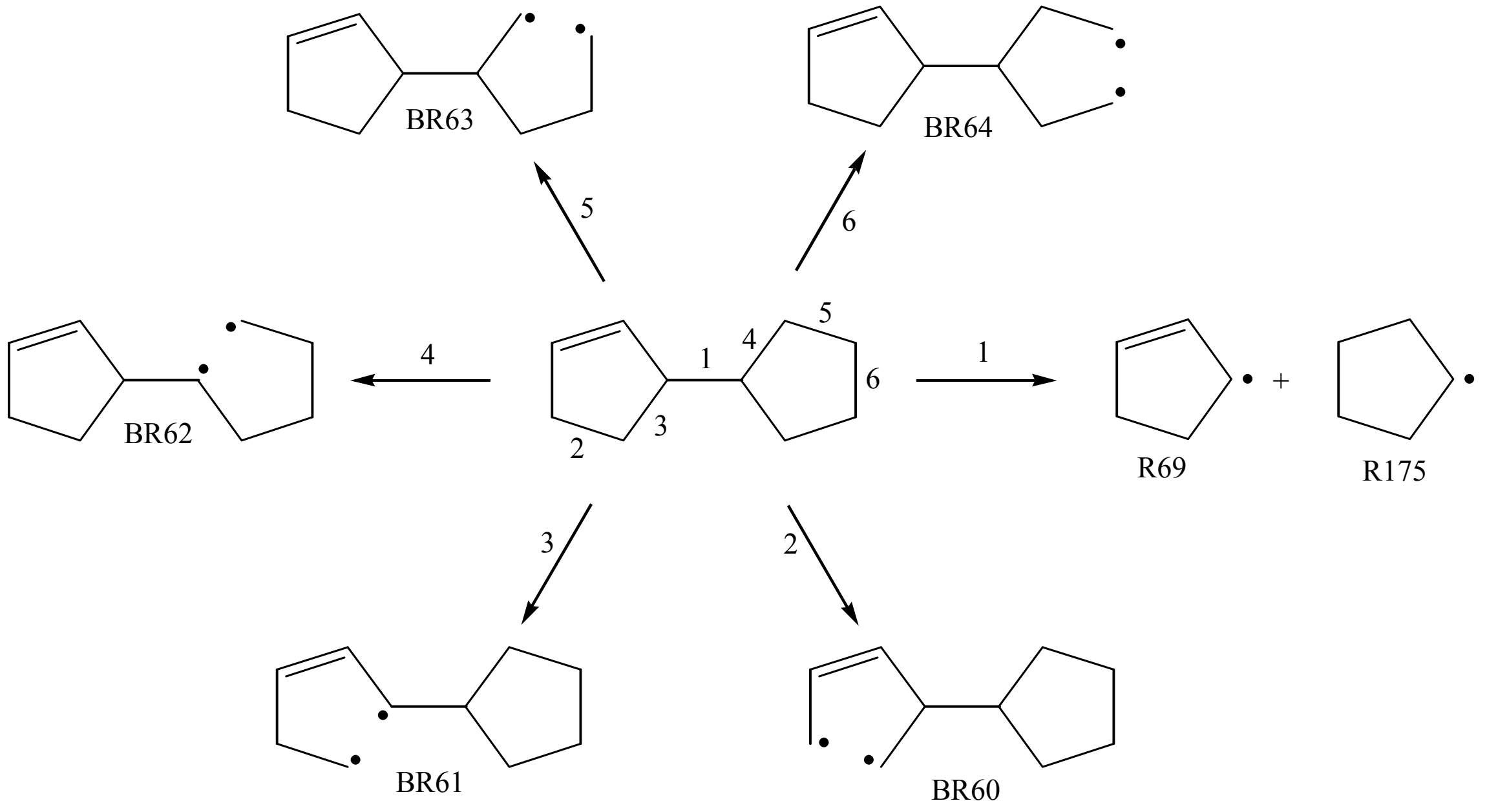

Figure 7: The 2 radicals and the 5 diradicals which can be obtained from 3-cyclopentyl-cyclopentene by unimolecular decomposition involving the breaking of a C-C bond.

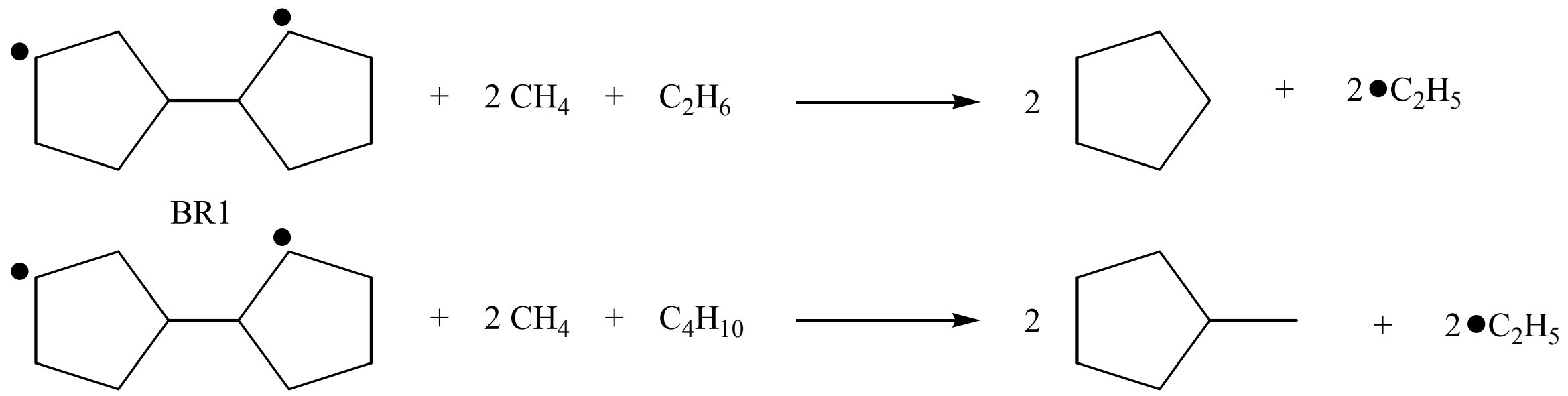

Figure 8: Two isodesmic reactions which were used for the estimation of the enthalpy of reaction of diradical BR1.

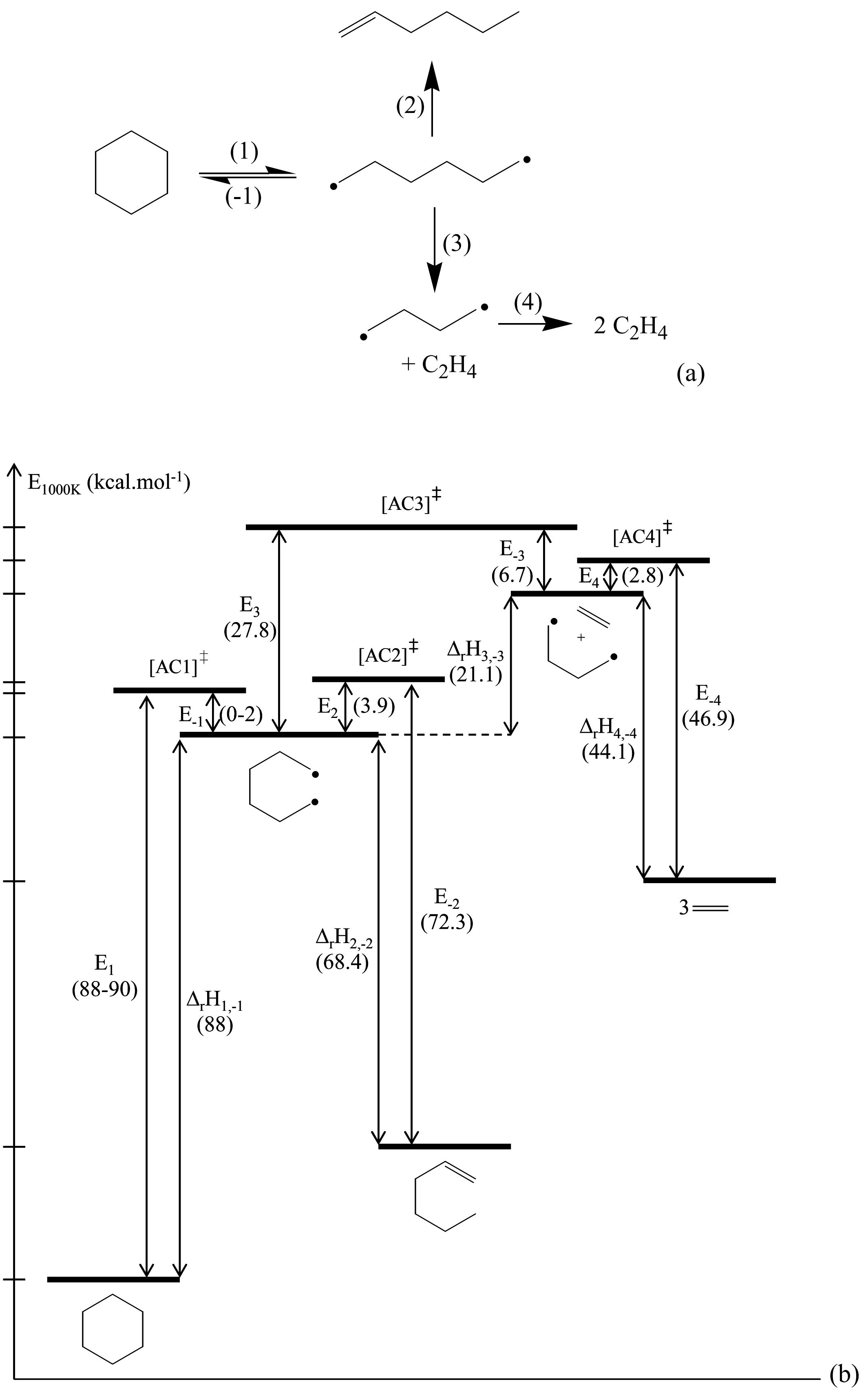

Figure 9: Fate of the diradical obtained by unimolecular initiation from cyclohexane (a) and energy levels associated to the reactions involved in the cyclohexane unimolecular initiation step (b).

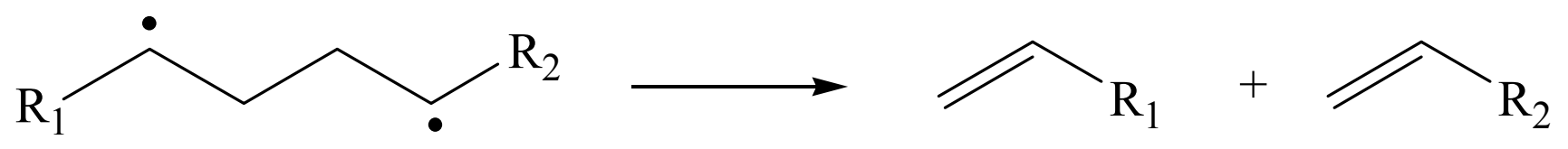

(a)

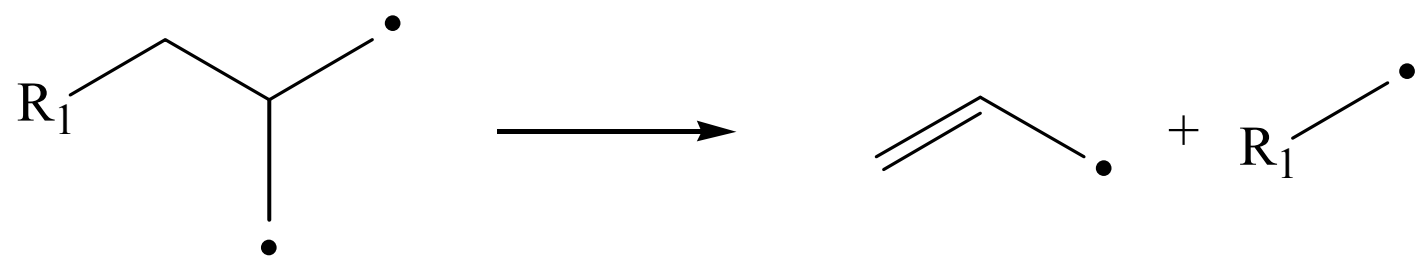

(b)

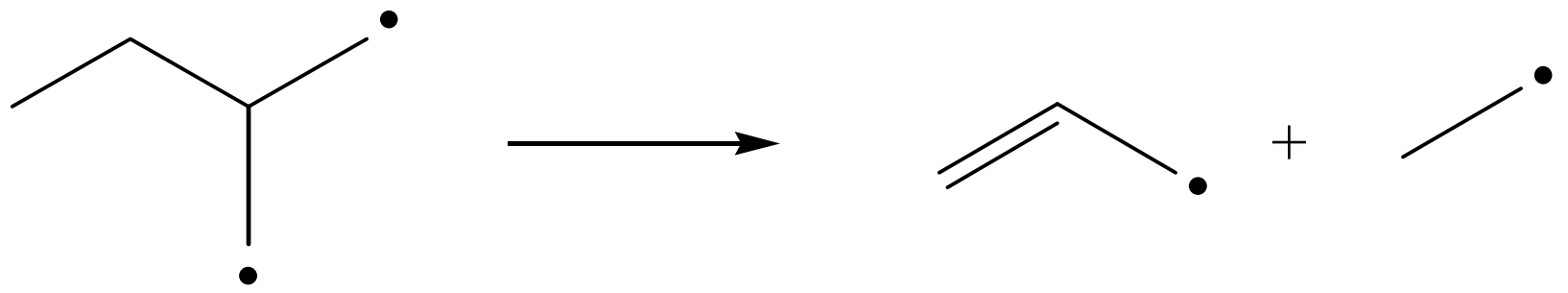

$\Delta_r H$=6.78 kcal.mol$^{-1}$

(c)

Figure 10: Scheme of β-scission decompositions involving the breaking of a Csp$^3$-Csp$^3$ bond in β−position of two radical centers (a, b) and model reaction used for the determination of the activation energy of reaction b (c).

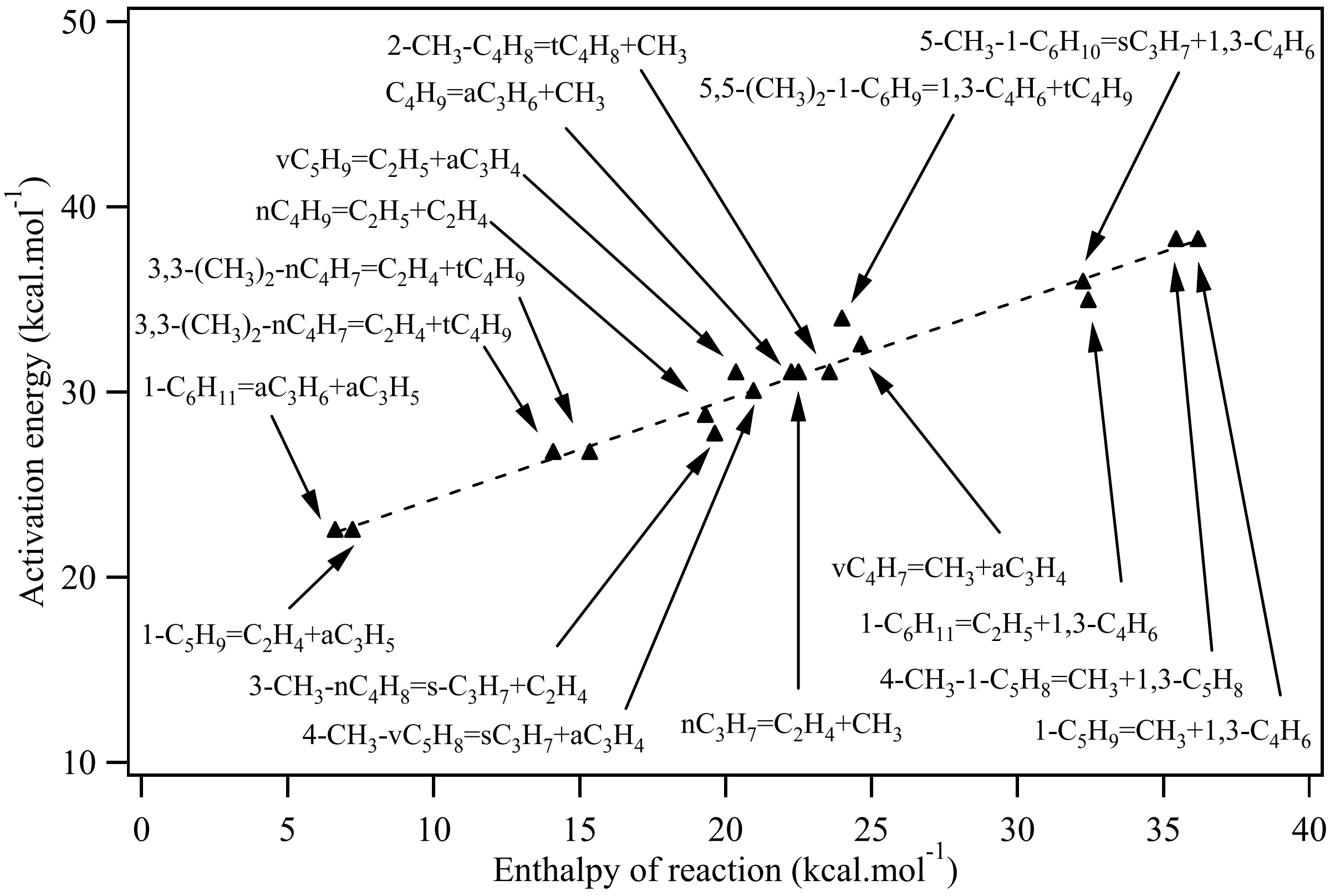

Figure 11: Activation energy of β-scission decomposition of acyclic compounds $Ea_{(acyclic\ \beta\text{-}scission)}$ vs. enthalpy of reaction $\Delta_rH_{(1000\ K)}$ (17).

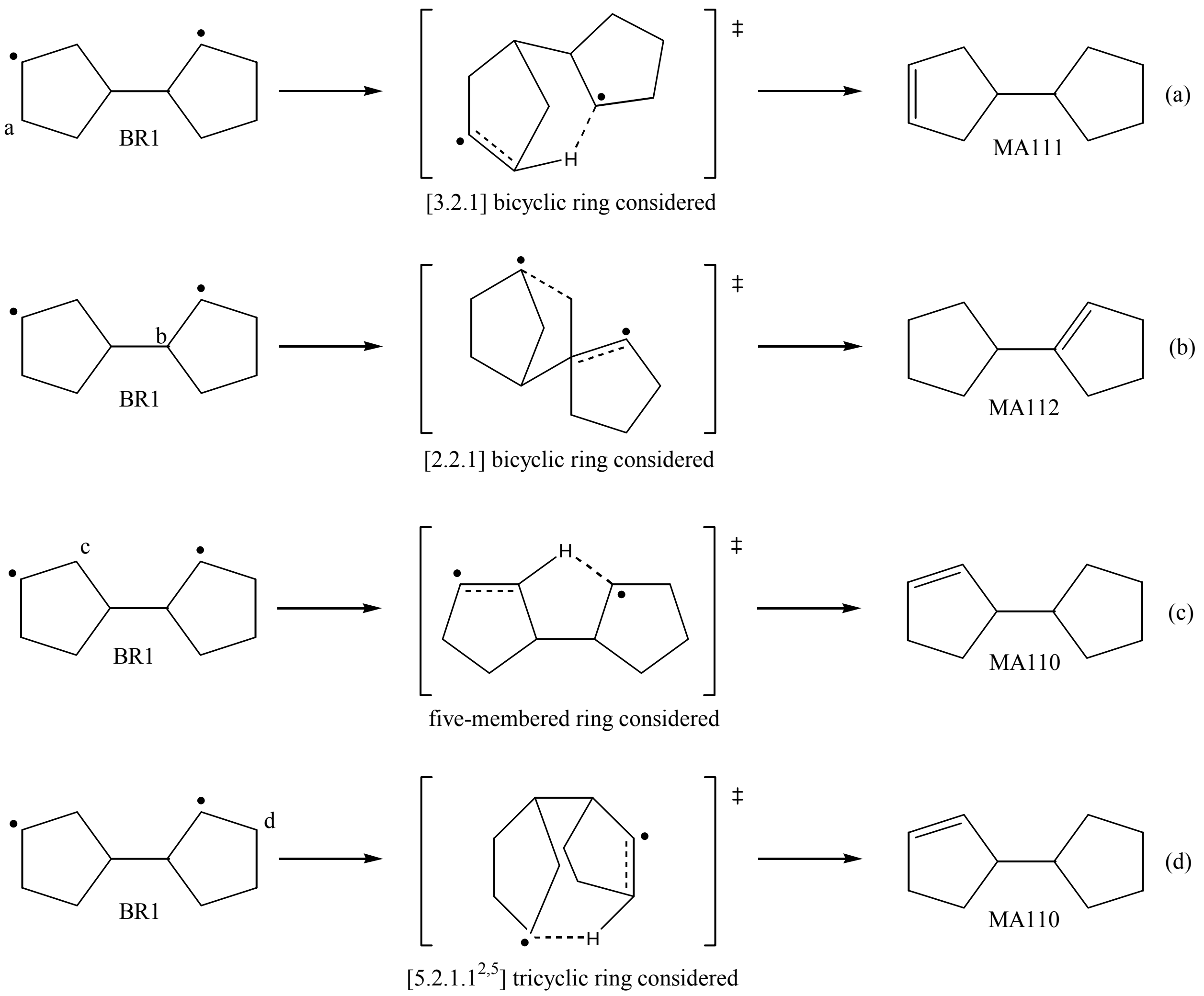

Figure 12: Structures assumed for the transition states of the isomerizations of Table 3.

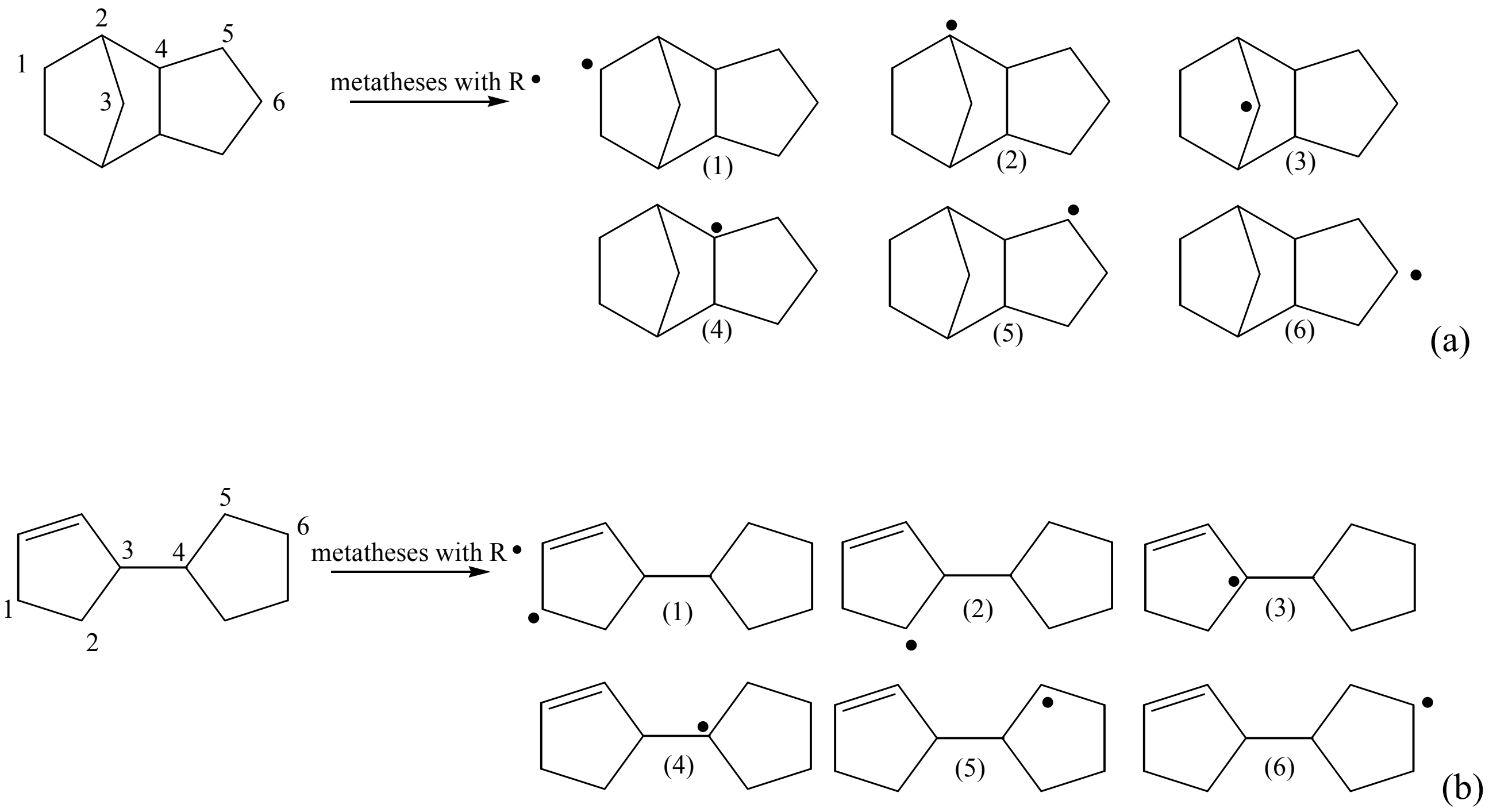

Figure 13: The 6 tricyclic free radicals obtained from metatheses with tricyclodecane (a) and the 6 bicyclic free radicals obtained from metatheses with 3-cyclopentyl-cyclopentene (b).

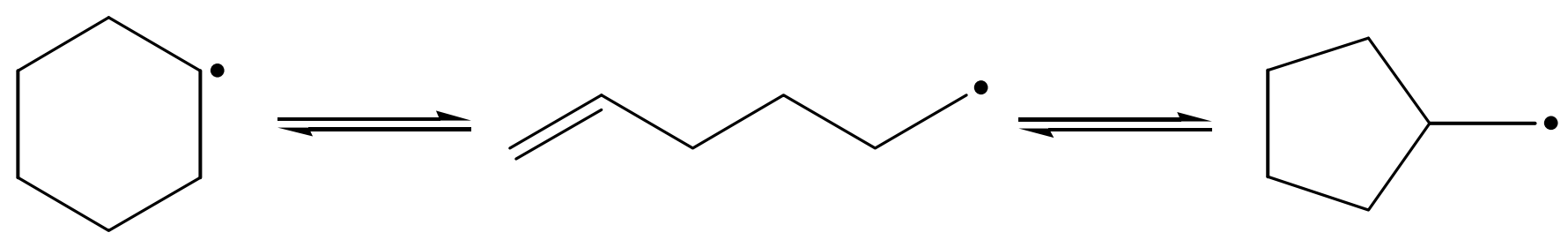

Figure 14: β-scission decomposition of the cyclohexyl radical and the two reverse reactions of intra-addition.

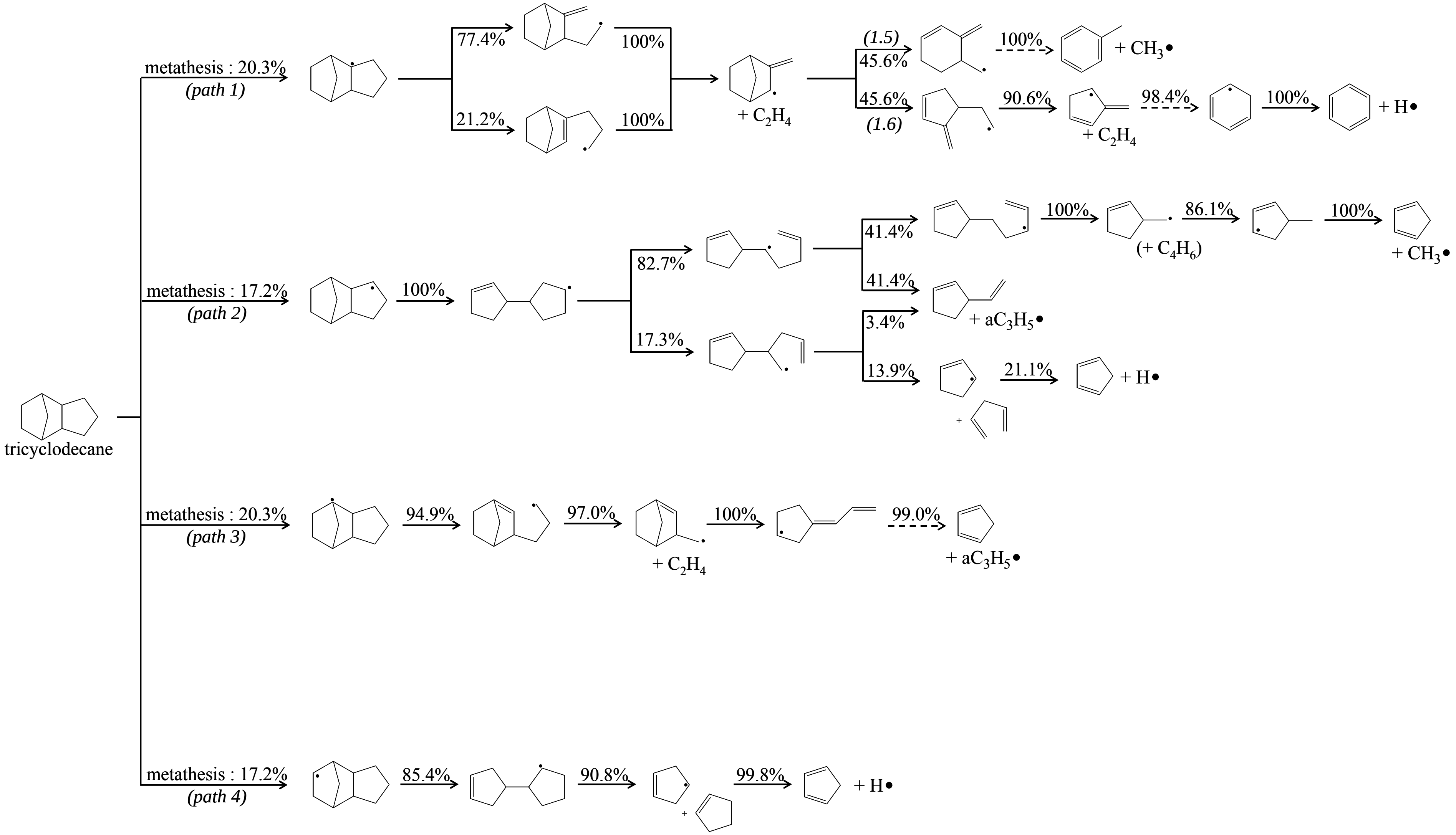

Figure 15: Major flows of consumption of tricyclodecane at a temperature of 933 K, a residence time of 1 s and a initial hydrocarbon mole fraction of 4% (corresponding to a conversion of 5.25%).

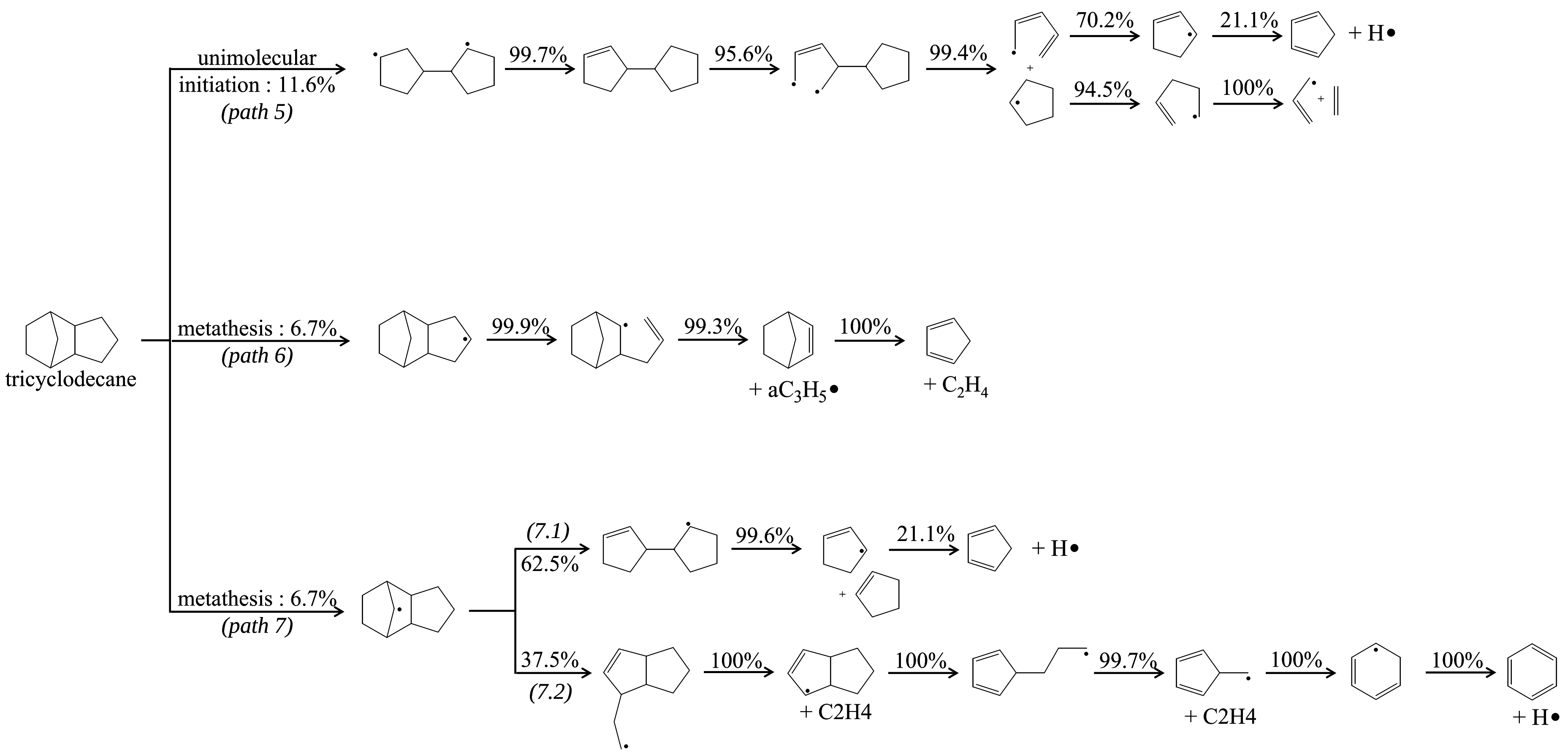

Figure 16: Minor flows of consumption of tricyclodecane at a temperature of 933 K, a residence time of 1 s and a initial hydrocarbon mole fraction of 4% (corresponding to a conversion of 5.25%).

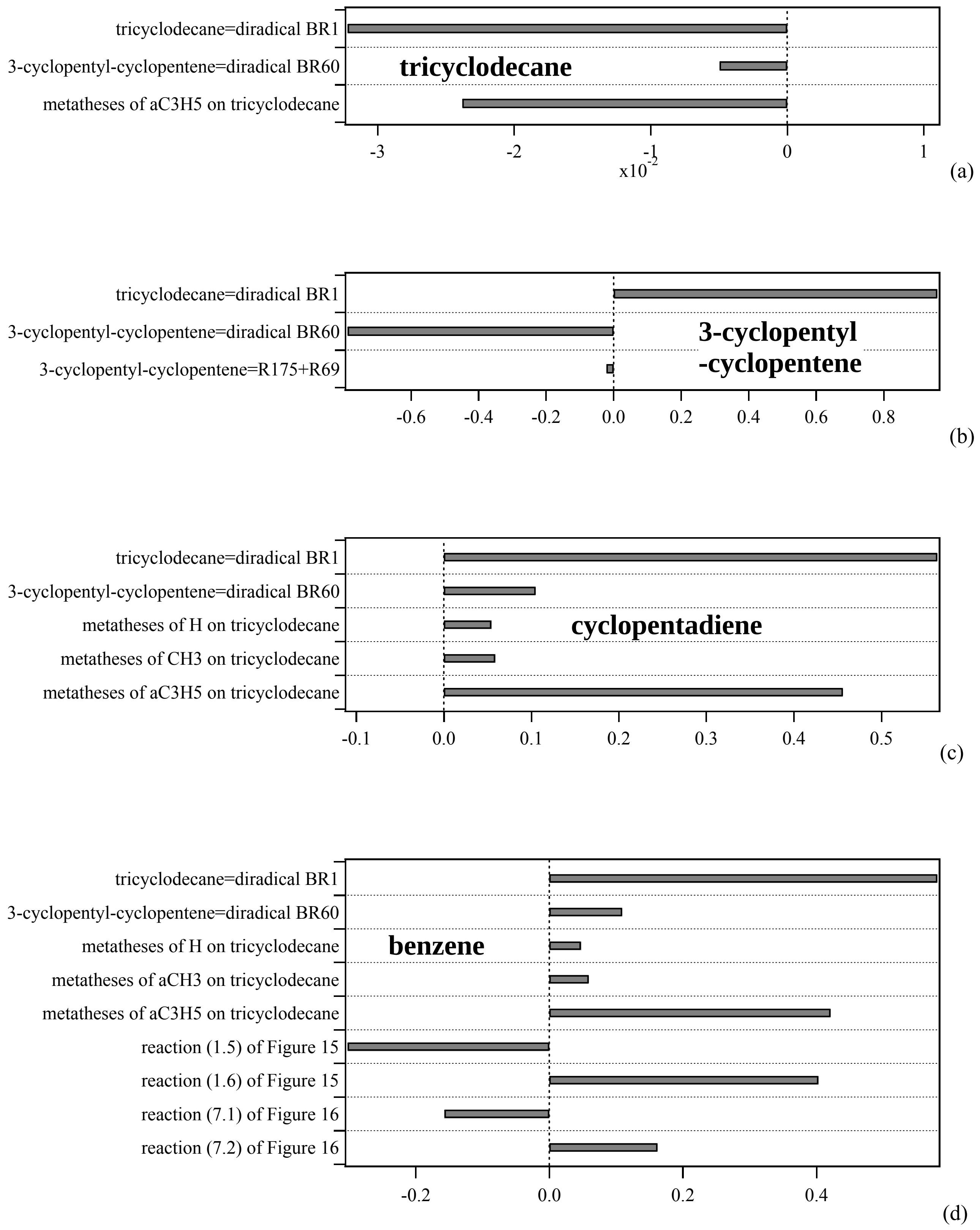

Figure 17: Sensitivity analyses for (a) tricyclodecane, (b) 3-cyclopentyl-cyclopentene, (c) cyclopentadiene and (d) benzene (T=933 K, τ=1 s).